\documentclass[10pt,twocolumn]{revtex4-1}
\usepackage{amsmath,amssymb,graphicx,bm,nccmath}
\usepackage{epsfig,subfigure}

\begin{document}
\author{Yamen Hamdouni}
\email{hamdouniyamen@gmail.com}
\affiliation{Department of physics, Faculty of Exact Sciences, Mentouri University, Constantine, Algeria}

\title{Decay rates and decoherence of an interstitial two-level spin impurity in a ferromagnetic   lattice}

\begin{abstract}
The decay rate of an interstitial  two-level spin impurity, located in the center of a unit cell of an anisotropic  ferromagnetic  lattice  subjected to an external magnetic field is derived. The impurity is coupled to nearest-neighbor  spins  through Heisenberg $XY$ interaction.  By mapping the lattice spin operators using the Holstein-Primakoff transformation, we establish the similarity with the Fano-Anderson model at low temperatures, and we calculate the retarded Green's function in one and two dimensions analytically for arbitrary coupling strength.  It is shown that the reduced density matrix of the impurity satisfies  an exact master equation in Lindblad form, from which the decay rate and the Lamb shift are deduced.  The evolution in  time of the latter together with  the excited state occupation probability  is investigated and its dependence on the applied magnetic field is discussed.   It is found that there exists a critical resonance-like value of the magnetic field around which the behavior of the decay rate and the density matrix changes drastically. The Markovian decay law, as given by the Fermi golden rule, does not hold in the weak-coupling regime unless the magnetic field is weak, typically less than the critical value. The weak-coupling regime is further treated perturbatively up to second order, and the obtained results are compared with the exact solution. We also discuss the Zeno regime of the dynamics,  where it is shown that at short times, the  effective decay rate is twice as small as the exact decay rate, and that when the impurity energy lies outside the lattice continuum, the measurement speeds up the decay of the survival probability.
\end{abstract}
\maketitle

\section{Introduction}

The complete description of the dynamics of (small) quantum systems should take into account the influence of the surrounding environment on the different features of their evolution. This  represents the basic concept behind the   theory of open quantum systems~\cite{petru}. As a  matter of fact,  many interesting phenomena  cannot be explained in a plausible way without the inclusion of the effect of the outer environment. The prominent examples that attracted much attention include  decoherence, dephasing and dissipation phenomena, to name a few ~\cite{zurek1, loss, zurek2}. Apart from their fundamental relevance in the development of quantum mechanics, these processes are of great importance in many applications, ranging from nuclear physics to quantum optics. 

Very often, the properties of the environment, which {\it a priori}, is characterized by   a large number of degrees of freedom, make it very difficult, if not impossible, to solve in  an exact manner the evolution equations. Fortunately, there exist systems of great practical relevance, for which the dynamics can be exactly solved. For instance, the Jaynes-Cummings model~\cite{jaynes}  represents one of the most popular and important paradigms that enabled the investigation of the dynamics of open quantum systems.  It has been widely used in many contexts and it is of great usefulness, both theoretically and experimentally.  
 Depending on whether the environment is of bosonic or  of fermionic   nature, many techniques have been proposed in order to eliminate the irrelevant environment degrees of freedom ~\cite{kha, coish, zhang, fazio, goan, sadi, bedoor, burg, ham1,ham2, ham3, paga}. Generally speaking, in the bosonic case, this  task is achieved  through the introduction of a spectral density for the system-environment coupling (usually of Lorentzian form), along with  the so-called  Born and Markovian approximations. The latter is widely used in, e.g., quantum optics, and is based on the assumptions that the characteristic time scale of the environment is much smaller that that associated with the central system. This leads to a loss of memory of the  system, which is generally associated with Markov processes. As a consequence, the reduced system density matrix is found to satisfy a master equation which is in the Lindblad form. The latter is characterized by  decay rates which are essentially positive and  time-independent.

However, the validity of the Markovian approximation is not justified in many systems that display features indicating strong  non-Markovian behavior. This is for example the case when the decay rates become negative implying that information flows back from the environment to the system; consequently, the memory effects should be taken into account even for weak coupling. Actually, the non-Markovian  dynamics of  quantum systems became over the last years one of the most interesting subjects in the theory of open quantum systems~\cite{diosi1,pilo1,pilo2}.  This is mainly due to the lack of an exact general non-Markovian master equation, in contrast to the known Lindblad form of the Markovian dynamics.  

The Fano-Anderson model describes a single discrete state or impurity that is coupled to a continuum of states. It was first introduced by Fano \cite{fano} and Anderson \cite{ande} to study magnetic impurities in metals. Notice that the impurity spin models are often met in the field of solid-state physics where the continuum may refer for example to an electron gas \cite{mahan,leggett}. This model, among other paradigms, has been the main tool in approaching various physical problems such as the spontaneous emission in dielectrics and photonic crystals, charge transfer  in one-dimensional semiconductors, the Bose-Einstein condensate, and the decay in Josephson junctions~ \cite{, john, nakazo,lambro, petro,tanaka,longhi1,segal,longhi2,zheng, zhang2,engel,shi,liu,longhi3,schmidt1,schmidt2,visuri,lena,mumford}. Recently, the general non-Markovian dynamics of open systems has been investigated via the use of Green's function for systems linearly coupled to thermal environments by Zhang {\it et al} \cite{,zhang2}. They used a model similar to the Fano-Anderson one, and showed how the exact master equation may be derived. In most investigations, the spectral densities of the environments studied  are defined over an infinite domain of mode frequencies where a frequency cut-off is introduced. Generally speaking, the exponential Markovian decay occurs  for weak system-environment coupling~\cite{tannoudji}. In Ref. \cite{john},  the authors report on  an oscillatory variation of the decay of the spontaneous emission of a two-level atom coupled to a radiation field whose spectrum possesses band gaps. The decay of the population on the excited state displays mostly non-Markovian dynamics for small detuning from the atomic resonant frequency. For large detuning, the decay becomes nearly Markovian (exponential); the same behavior has been reported in \cite{lambro}. 

In this paper, we focus on the study of the non-Markovian dynamics of a central spin impurity that is coupled to a ferromagnetic spin lattice. The latter presents periodic properties~\cite{majlis,kittel} that fix in a unique manner the spectral density. It should be stressed that the spin degrees of freedom are the most suitable candidates towards   the implementation of new quantum technologies~\cite{loss2,loss3}.  One can, for instance, profit from their properties to
implement the proposed quantum algorithms \cite{nielsen, shor, ekert, cirac, mosca, chuang}. In this work, we shall be mainly interested in the decay rates, whose properties determine the way the reduced density matrix behaves in time.

The paper is organized as follows. In Sec.~\ref{sec2} we introduce the total Hamiltonian of the composite system. Then, through the Holstein-Primakoff transformation,  we use the spin-wave theory to establish the connection with the Fano-Anderson model. Section \ref{sec3} deals with the study  of the dynamics of the impurity at zero temperature, where  analytical results for one and two dimensions are presented, and the evolution of the decay rate and the occupation probability is discussed. Section \ref{sec4} is devoted to the  study  of the weak-coupling regime where we use the second-order perturbation theory to derive the master equation for the reduced density matrix, and we compare the results with the exact solutions. There, the short-time evolution is discussed in more details. In Sec.~\ref{sec5} we investigate the quantum Zeno effect. We end  the paper with a brief conclusion.

\section{Model \label{sec2}}
\subsection{System Hamiltonian}
Consider a  two-level localized spin impurity that is immersed in a ferromagnetic spin lattice in $d$ dimensions. The impurity is dealt with as a central
open system, while the lattice plays the role of the spin bath. The total model Hamiltonian $H$ is
given by the sum of three terms: the  free Hamiltonian of the central system which we designate by $H_S$,  the Hamiltonian of the lattice $H_B$, and the interaction Hamiltonian $H_{SB}$ describing 
the coupling of the impurity  to the spins  of the lattice. Therefore, we can write:
\begin{equation}
 H=H_S+H_B+H_{SB}.
\end{equation}
The free Hamiltonian of the two-level system may be expressed in terms of the usual Pauli matrices as:
\begin{equation}
 H_S=\omega_0 \sigma_+\sigma_-,
\end{equation}
where $\omega_0$ is the energy gap between the ground state and the excited state of the impurity. Note that the formalism we  use  applies  as well to the 
case of a qubit where the free Hamiltonian is written as $H_S=(\omega_0/2) \sigma_z$,  $\omega_0$ being proportional to the strength of the local  magnetic field applied to the qubit. 
 
The   lattice is subject to the effect of a homogeneous magnetic field, applied along the $z$-direction, the strength of which is denoted by $h$. The Hamiltonian describing the lattice reads:
\begin{equation}
   H_B=-\sum\limits_{\langle i,j\rangle}( J_{i,j}^x S_i^xS_j^x+J_{i,j}^y S_i^yS_j^y+J_{i,j}^z S_i^zS_j^z)-h\sum_j S_j^z,
\end{equation}
where  $S_i^x$,  $S_i^y$ and  $S_i^z$ represent the components of the spin operator of the spin of magnitude $S$ located at site $i$. In the above equation, the summation is performed 
with respect to all pairs of spins. The parameters $J_{i,j}^x$, $J_{i,j}^y$ and $J_{i,j}^z$ denote the coupling constants which are all positive. We assume that the lattice is of $XXZ$ type, and that  each spin   interacts only with its nearest neighbors, whose  number  is denoted from here on by $\eta$ (the coordination number). 

Under the above assumptions, the lattice Hamiltonian can be written as
\begin{equation}
 H_B=-J\sum\limits_{j{\bm {\delta}}}( S^x_j S^x_{j+\bm{\delta}}+ S^y_j S^y_{j+\bm {\delta}}+\gamma_z S^z_j S^z_{j+\bm{\delta}})-h\sum_j S_j^z,
\end{equation}
where  $\bm \delta$ designates the $d$-dimensional vectors joining each spin at a given site  to its nearest neighbor spins, and $J=J^x_{i,i+\bm \delta}=J^y_{i,i+\bm \delta}$  denotes the coupling constant restricted to these  neighbors, whereas $\gamma_z=J^z_{i,i+\bm \delta}/J$ is  the anisotropy parameter which satisfies $\gamma_z \ge1$. This easy-axis condition  ensures that the lattice is in the ferromagnetic phase, where  the ground state  is the one in which all the spins are directed along the $z$-direction.
At this stage, it is useful to introduce the raising and lowering operators $S_j^\pm=S_j^x\pm i S_j^y$, which enables us to rewrite the lattice Hamiltonian in the form:
\begin{equation}\label{hami} 
 H_B=-\frac{J}{2}\sum\limits_{j{\bm {\delta}}}\Bigl[ ( S^+_j S^-_{j+\bm {\delta}}+S^-_j S^+_{j+\bm {\delta}})+ 2\gamma_z S^z_j S^z_{j+\bm{\delta}})\Bigr]-h\sum_j S_j^z.
\end{equation}

We further assume that the  coupling between the spin impurity and the lattice is of Heisenberg $XY$ type, whose Hamiltonian is given explicitly by the formula:
\begin{equation}
 H_{SB}=\sum_j (g_j \sigma_-S^+_j+g_j^* \sigma_+S^-_j),
\end{equation}
where  $g_j$ denotes the coupling constant of the central system to the spin located at  site $j$; for the sake of generality, we assume it to be complex-valued.
\subsection{Spin-wave formulation}

The properties of ferromagnets at low temperatures can be investigated by means of the spin-wave theory, where the concept of the magnon naturally arises as the analog
 of the photon in electromagnetic radiations, and of the phonon for the lattice vibrations. Generally speaking, magnons are  ground state excitations that propagate through
 the spin lattice, as a result of thermal or quantum perturbations.  The standard method in spin-wave theory consists in using suitable transformations that map the spin 
 operators to bosonic operators. In this work we use the Holstein-Primakoff transformation which proved to be very convenient in solving such problems. Recall that the 
 prescription employed in  the  Holstein-Primakoff transformation resides in the following identities~\cite{holstein}:
 
\begin{eqnarray}
  S^-_j&=&\sqrt{2S}\sqrt{1-\tfrac{b_j^\dag b_j}{2S}} b_j, \quad   S^+_j=\sqrt{2S}b^\dag_j\sqrt{1-\tfrac{b_j^\dag b_j}{2S}},\label{hols}\\
  && S_j^z=S-b_j^\dag b_j,
\end{eqnarray}
where $b_j$ are bosonic operators that satisfy  $[b_j, b_{j'}^\dag]=\delta_{jj'}$. 

At low temperatures, the mean number of magnons is very small; therefore, by expanding the square root in Eq.~(\ref{hols}) in a Taylor series and keeping only bilinear terms  in $b_j$ and $b_j^\dagger$ in the Hamiltonian of the lattice, it follows that:
\begin{eqnarray}
 H_B&=&-JS\sum\limits_{j{\bm {\delta}}}\Bigl[ (b_j^\dag b_{j+\bm{\delta}}+b_j  b_{j+\bm{\delta}}^\dag)+2\gamma_z (b_j^\dag b_{j}+b_{j+\bm{\delta}}^\dag b_{j+\bm{\delta}})\Bigr]
\nonumber \\ &+& h\sum_j b_j^\dag b_j-hNS-J\eta\gamma_z N S^2,
\end{eqnarray}
where $N$ is the number of sites or spins in the lattice. Next, we  Fourier transform the bosonic operators $b_j$ as follows:
\begin{equation}
 b_j=\frac{1}{\sqrt{N}}\sum_{\vec k} e^{i \vec k \vec r_j} a_{\vec k}, \quad
 a_{\vec k} =\frac{1}{\sqrt{N}}\sum_j e^{-i \vec k \vec r_j} b_j, \label{four}
\end{equation}
 where  $\vec r_j$ designates the $d$-dimensional real-space vector that determines the  position of the spin at site $j$ of the lattice. 
It can easily be verified that the operators $a_{\vec k}$ satisfy $[a_{\vec k}, a_{\vec k'}^\dag]=\delta_{\vec k \vec k'}$. By virtue of Eq.(\ref{four}), the lattice Hamiltonian is written as:
\begin{eqnarray}
 H_B &=&\sum_{\vec k}\bigl(h+2J\gamma_z\eta S-J \eta S 2\tau_{\vec k}\bigr) a_{\vec k}^\dag a_{\vec k}
-hNS \nonumber \\
 &-&J\eta S/2\sum_{\vec k} \tau_{\vec k} -J\eta\gamma_z N S^2 \label{hamo},
\end{eqnarray}
where
\begin{equation}
 \tau_{ \vec k}=\frac{1}{\eta}\sum_{\vec {\bm \delta}} e^{i \vec k \vec{\bm \delta}}
\end{equation}
is the lattice structure factor. Hence, we deduce that the  dispersion relation is given by:
\begin{equation}
 \Omega_{\vec k}=h-2J\eta S (\tau_{\vec k}-\gamma_z).
\end{equation}
This implies that the spectrum of the lattice  corresponds to the energy domain  $\Omega_{\vec k} \in [\Omega_{\rm min},\Omega_{\rm max}]$, with
$\Omega_{\rm min}=h+2J\eta S (\gamma_z-1)$ and $\Omega_{\rm max}=h+2J\eta S (1+\gamma_z)$ (we set $\hbar=1$).

In a similar way, it can be shown that the interaction Hamiltonian $H_{SB}$ may be expressed in terms of the bosons operators as:
\begin{eqnarray}
 H_{SB}&=&\sum_{\vec k} (g_{\vec k} \sigma_{-}a_{\vec k}^++g_{\vec k}^* \sigma_+a_{\vec k}),
 \end{eqnarray}
where the new coupling constant $g_{\vec{k}}$ is defined through the expression:
\begin{equation}
 g_{\vec k}=\sqrt{\frac{2S}{N}}\sum_j g_j e^{i\vec{k}\vec{r_j}}.\label{coup}
\end{equation}
 Clearly, the  latter form of the coupling constant depends on the position of the impurity, as well as on  dimension and  type of  the lattice.
 \section{Exact dynamics  at zero temperature ($T=0$)\label{sec3}}
The evolution in time of the state of the spin impurity at zero temperature can be exactly derived.
We begin with the one-dimensional lattice where we present the main calculations and procedures, which are of general applicability and hold at higher dimensions; afterwards, we deal with the two-dimensional case where we show that the dynamics can also be investigated analytically.
\subsection{One-dimensional lattice}
The dynamics in one dimension bears a particular importance as certain interesting  features arise, which are absent in two and three dimensions.  The lattice in this case is a one-dimensional  array of spins, the length of which is given by $(N-1)\delta$, where $\delta$ represents  the distance separating two adjacent sites. The wave vector has only one component $k$, and the first Brillouin zone corresponds to the interval $-\pi/\delta \le k \le \pi/\delta$. We assume that the impurity lies in the middle between two lattice spins, and that it interacts only with these two neighbors, with coupling constant $g$. The nonlocal coupling constant $g_{\vec k}$ is thus given by:
\begin{equation}
 g_{\vec k}=2g\sqrt{\frac{2S}{N}} \cos\left(\frac{\delta k}{2}\right). \label{coup1}
\end{equation}
 Next, we introduce the retarded Green's function of the impurity which is given by:
\begin{equation}
G_{\rm ret}(\epsilon)=\frac{1}{ \epsilon-\omega_0-\Sigma_{\rm ret}( \epsilon+i\nu)},\label{green}
\end{equation}
where $\Sigma_{\rm ret}( \epsilon+i\nu)$ is the retarded self-energy, namely:
\begin{equation}
\Sigma_{\rm ret}( \epsilon+i \nu)=\sum_k\frac{|g_{\vec k}|^2}{\epsilon-\Omega_{\vec{k}}+i\nu},
\end{equation}
with $\nu$ being  an infinitesimal positive quantity.  It should be stressed that due to the $XY$ coupling of the impurity to the lattice, the only self-energy diagram after bosonization is the one obtained in the second-order  expansion of the $S$-matrix with respect to the coupling constants $g_{\vec k}$, in analogy with the Fano-Anderson model. Hence the Green's function~(\ref{green}) is actually an exact one that is obtained by summing all the diagrams,  which amounts to writing~\cite{mahan}:
\begin{eqnarray}
 G(\epsilon)&=&G_0(\epsilon)+G_0(\epsilon)\Sigma(\epsilon)\left[G_0(\epsilon)+G_0^2(\epsilon)\Sigma(\epsilon)+\cdots\right]\nonumber \\
 &=&\frac{G_0(\epsilon)}{1-G_0 (\epsilon) \Sigma(\epsilon)},
\end{eqnarray}
 where $G_0(\epsilon)=(\epsilon-\omega_0)^{-1}$ is the Green's function of the free impurity, and $\Sigma(\epsilon)$ is the self-energy. 

In the limit $N\to \infty$, the number of modes becomes very large, and hence the spectrum of the lattice turns into a continuum of states, whose energies are bounded between $h+4JS(\gamma_z-1)$ and $h+4JS(\gamma_z+1)$. In this limit, in $d$ dimensions, the sum with respect to the wave vector $\vec k$ may be replaced with an integral over the first Brillouin zone (FBZ)  according  to the  rule:
\begin{equation}
 \frac{1}{N}\sum_{\vec k}\to \frac{v_d}{(2\pi)^{d}} \int_{FBZ} d\vec{k},
\end{equation}
where  $v_d$ is the volume of a unit cell. 
Consequently, by virtue of Eq.~(\ref{coup1}), the retarded self-energy can be expressed as
\begin{equation}
\Sigma_{\rm ret}(\epsilon+i\nu)=\frac{4g^2 S}{\pi}\int_{-\pi}^{\pi}\frac{\cos^2(k/2) dk}{\epsilon-h-4JS\gamma_z+4JS\cos(k)+i\nu}.
\end{equation}
 By the change of variable $z=e^{ik}$, the above integral is transformed into a contour integral around the unit circle, thereby the theorem of residues can be applied. It follows that the real part of the self-energy reads:
\begin{widetext}
\begin{eqnarray}
{\rm Re}\ \Sigma_{\rm ret}(\epsilon+i\nu)=\begin{cases}
           \frac{g^2}{J}\left(1-\sqrt{\frac{\epsilon-h-4JS\gamma_z-4JS}{\epsilon-h-4JS\gamma_z+4JS}}\right) & \text{for}  \quad 4JS\leq|\epsilon-h-4JS\gamma_z| \\
           \frac{g^2}{J}  &  \text{for}   \quad 4JS>|\epsilon-h-4JS\gamma_z| .\label{onerel}
\end{cases}
\end{eqnarray}
\end{widetext}
On the other hand, the imaginary part can be calculated when $\nu\to 0^+$ as:
\begin{widetext}
\begin{equation}
{\rm Im}\ \Sigma_{\rm ret}(\epsilon+i0^+)=-\frac{g^2}{J}\left(\frac{h-\epsilon+4JS(1+\gamma_z)}{\sqrt{(4JS)^2-(\epsilon-h-4JS\gamma_z)^2}}\right) \quad \text{for}  \quad 4JS>|\epsilon-h-4JS\gamma_z|,
\end{equation}
\end{widetext}
and  ${\rm Im}\ \Sigma_{\rm ret}(\epsilon+i0^+)=0$ for $4JS<|\epsilon-h-4JS\gamma_z|$. This actually reflects the fact that   the imaginary part of the retarded self-energy vanishes outside the continuum, namely ${\rm Im}\ \Sigma_{\rm ret}(\epsilon+i\nu)=0$ for $\epsilon>\Omega_{\rm max} $ or $\epsilon<\Omega_{\rm min}$. 

\begin{figure}[htb]
{\centering{\resizebox*{0.38\textwidth}{!}{\includegraphics{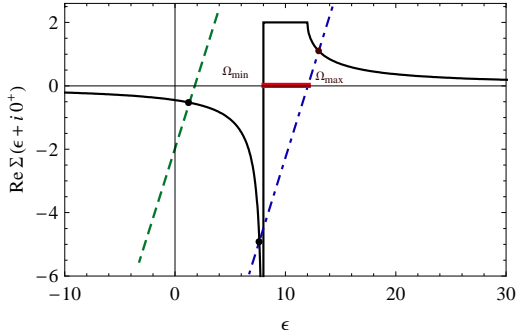}}}
\caption{\label{figure1} The real part of the self-energy in one dimension  (black solid line) for $J=0.5$, $g=1$, $h=8$, $S=1$, and $\gamma_z=1$. The intersection points  correspond to the solutions of the equation ${\rm Re}\ \Sigma_{\rm ret}(\epsilon+i0^+)=\epsilon-\omega_0$ for $\omega_0=2 $ (green dashed line) and $\omega_0=12$ (blue dot-dashed line); these solutions are interpreted in the main text as localized bound states outside the continuum. }}
\end{figure}

The fundamental property exhibited by the retarded Green's function is that  the spectral density $A(\epsilon)$ is related to the former through the identity
\begin{equation} 
A(\epsilon)=-2 \ {\rm Im} \ G_{\rm ret}(\epsilon).
\end{equation}
The spectral density represents essentially the probability that the impurity has energy $\epsilon$, as a result of its coupling to the lattice. Inside the continuum, it can be expressed as:
  
\begin{equation} \medmath{
A(\epsilon)=- \frac{2\ {\rm Im}\ \Sigma_{\rm ret}(\epsilon+i0^+)}{\left(\epsilon-\omega_0-{\rm Re}\ \Sigma_{\rm ret}(\epsilon+i0^+)\right)^2+\left({\rm Im}\ \Sigma_{\rm ret}(\epsilon+i0^+)\right)^2}}. 
\end{equation}
Outside the continuum, that is when ${\rm Im}\ \Sigma_{\rm ret}(\epsilon+i0^+)\to 0$, the spectral density reduces to
\begin{eqnarray}
A(\epsilon)&=&2\pi \delta\left(\epsilon-\omega_0-{\rm Re}\ \Sigma_{\rm ret}(\epsilon+i0^+)\right)\nonumber\\
&=& 2\pi \sum_j \frac{\delta(\epsilon-\epsilon_j)}{1-\frac{d}{d\epsilon}{\rm Re}\ \Sigma_{\rm ret}(\epsilon+i0^+)|_{\epsilon_j}},
\end{eqnarray}
where $\epsilon_j$ are the solutions of the equation $\epsilon-\omega_0-{\rm Re}\ \Sigma_{\rm ret}(\epsilon+i0^+)=0$. They may be interpreted as the energies corresponding to localized bound states  outside the continuum. Thus, these states are determined in the one-dimensional case by solving the equation:
\begin{equation}
 \epsilon-\omega_0= \frac{g^2}{J}\left(1-\sqrt{\frac{\epsilon-h-4JS\gamma_z-4JS}{\epsilon-h-4JS\gamma_z+4JS}}\right)
\end{equation}
outside the continuum, which  can be carried out numerically; however to gain more insight into the existence of the bound states, it is convenient to discuss the solutions of the latter equation  graphically as  displayed in Fig.~\ref{figure1}. It can be shown that there exists always at least one solution no matter what the values of the model parameters are. This is due to the fact that the real part of the self-energy diverges at the lower edge of the spectrum, see Eq.~(\ref{onerel}). More precisely,  the impurity exhibits two bound states whenever  $\omega_0> h+4JS(1+\gamma_z)-g^2/J$, otherwise there exists only one bound state. Indeed, from Fig.~\ref{figure1}, we see that  the  value of $\omega_0$ for which the line $y=\epsilon-\omega_0$ passes through the point $(h+4JS(1+\gamma_z),g^2/J)$ is given by $ h+4JS(1+\gamma_z)-g^2/J$; all the values of $\omega_0$ exceeding the latter lead to two intersection points outside the continuum.

The retarded Green's function in the time domain is given by the Fourier  transform of the spectral density, that is:
\begin{equation} 
G_{\rm ret}(t)=-i\int\limits_{-\infty}^\infty \frac{d\epsilon}{2\pi} A(\epsilon) e^{-i\epsilon t}=-i \phi_+(t),
\end{equation}
where $\phi_+(t)$ is the wave function describing the excited state of the impurity. The integration runs over the full real line to ensure that all the poles of the Green's function outside the continuum are taken into account. In the present model, the range of integration is  determined by the energies of the  bound states, along with  the continuum; at the bound states the spectral density is given by a delta function. Hence:
\begin{widetext}
\begin{eqnarray}
 \phi_+(t)=\sum_j \frac{e^{-i\epsilon_j t}}{1-\frac{d}{d\epsilon}{\rm Re}\ \Sigma_{\rm ret}(\epsilon+i0^+)|_{\epsilon_j}}  -\frac{1}{\pi}\int\limits_{\Omega_{\rm min}}^{\Omega_{\rm max}} \frac{\ {\rm Im}\ \Sigma_{\rm ret}(\epsilon+i0^+)e^{-i\epsilon t} d\epsilon}{\left(\epsilon-\omega_0-{\rm Re}\ \Sigma_{\rm ret}(\epsilon+i0^+)\right)^2+\left({\rm Im}\ \Sigma_{\rm ret}(\epsilon+i0^+)\right)^2}. \label{ampl}
\end{eqnarray}\end{widetext}

\begin{figure}[htb]
{\centering\subfigure[]{\resizebox*{0.38\textwidth}{!}{\includegraphics{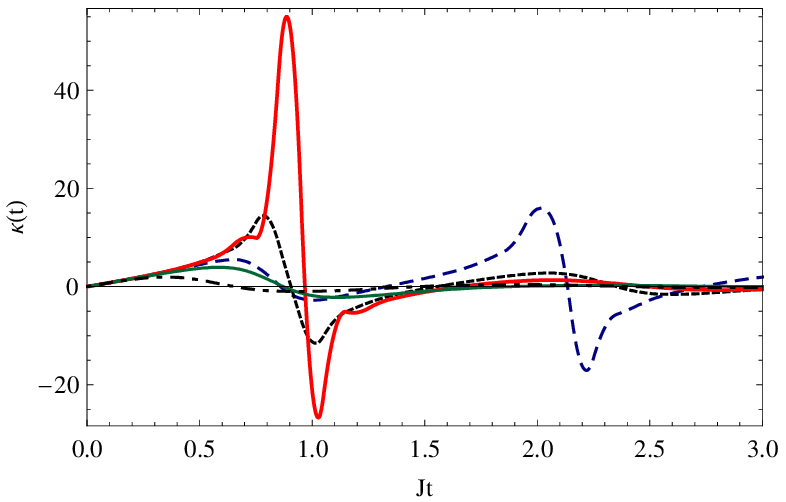}}}
\subfigure[]{
\resizebox*{0.38\textwidth}{!}{\includegraphics{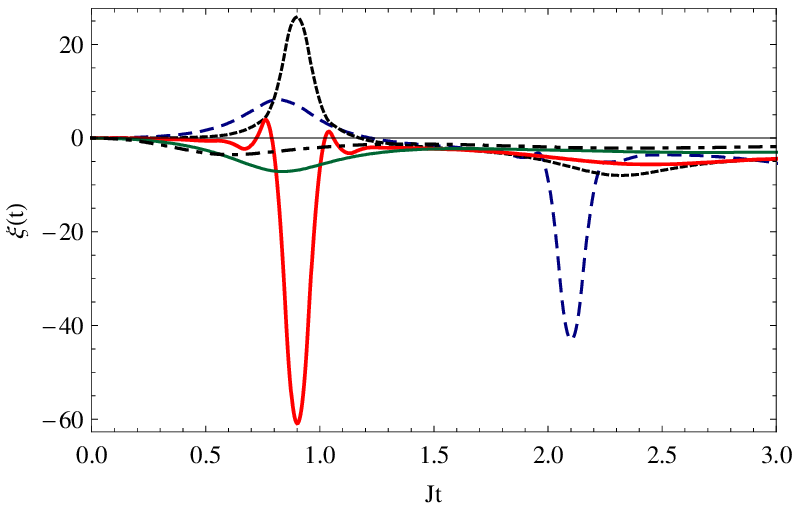}}}
\par}
\caption{\label{figure2}  (a) The decay rate $\kappa$, and (b) the Lamb shift $\xi$  for strong coupling as a function of the time for different values of the strength of the magnetic field: $h=0.1J$ (blue dashed lines), $h=J$ (black dotted lines), $h=1.5J$ (red thick solid lines), $h=3J$ (green thin solid lines), and $h=4J$ (black dot-dashed lines); other parameters are: $\omega_0=3J$, $S=1$, $g=J$, and $\gamma_z=1$.}
\end{figure}

\begin{figure}[htb]
{\centering\subfigure[]{\resizebox*{0.38\textwidth}{!}{\includegraphics{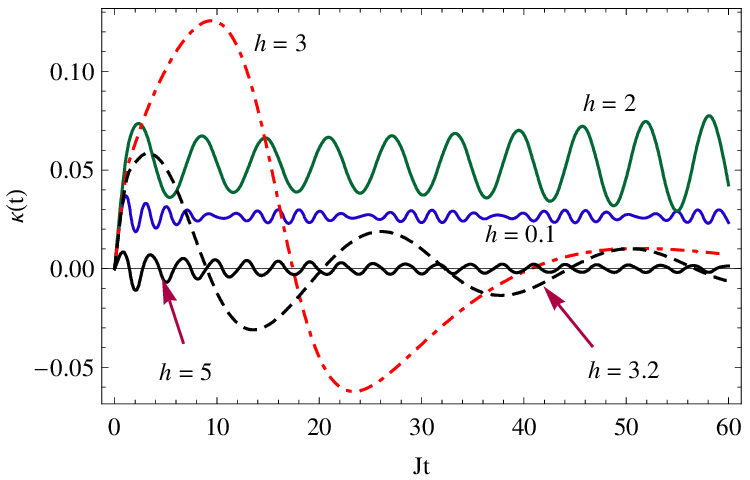}}}
\subfigure[]{
\resizebox*{0.38\textwidth}{!}{\includegraphics{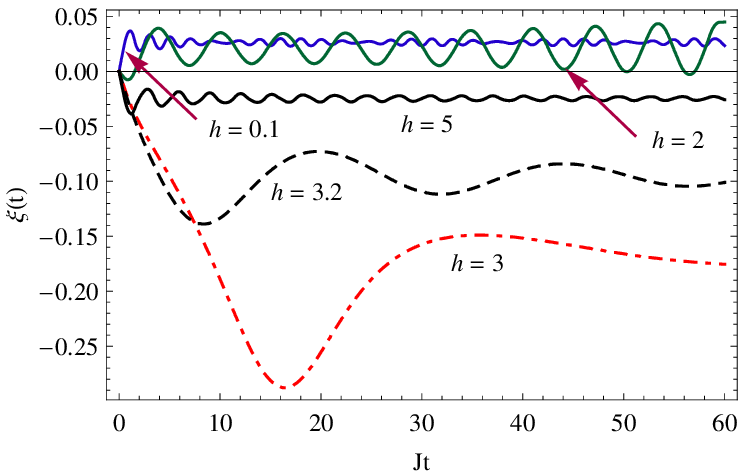}}}
\par}
\caption{\label{figure3} (a) The decay rate $\kappa$, and (a) the Lamb shift $\xi$  for weak coupling as a function of the time for different values of the strength of the magnetic field (in units of $J$): $h=0.1$ (blue solid lines), $h=2$ (green solid lines), $h=3$ (red dot-dashed lines), $h=3.2$ (black dashed lines), and $h=5$ (black solid lines); other parameters are: $g=0.1J$, $\omega_0=3J$, $S=1$, and $\gamma_z=1$.}
\end{figure}

Suppose that the initial state of the system is given by the pure state $(\alpha_+ |+\rangle+\alpha_- |-\rangle)|\mathcal G\rangle$
where $|\mathcal G\rangle=\otimes_k|0\rangle_k$ is the ground state of the lattice at $T=0$.  Then because of  the form of the interaction Hamiltonian $H_{SB}$, the state of the impurity evolves to the mixed one:
\begin{equation}
\rho(t)=\begin{pmatrix} |\alpha_+|^2 |\phi_+(t)|^2 &&  \alpha_-^* \alpha_+ \phi_+(t) \\
\alpha_- \alpha_+^* \phi_+(t)^* && 1-|\alpha_+|^2|\phi_+(t)|^2 
\end{pmatrix}\label{genden}.
\end{equation}
Differentiating both sides of Eq.~(\ref{genden})  with respect to time, and using the properties of the Pauli matrices, it can be shown that the above  density matrix  satisfies the exact master equation
\begin{eqnarray}
 \frac{d\rho(t)}{dt}&=&-i[(\omega_0+\xi(t)/2)\sigma_+\sigma_-,\rho(t)] \nonumber \\ &+&\kappa(t)\left(\sigma_{-}\rho(t)\sigma_{+}-\frac{1}{2}\{\sigma_{+}\sigma_{-},\rho(t)\}\right), \label{masteq}
\end{eqnarray}
where:
\begin{eqnarray}
\kappa(t)&=&-2 {\rm Re} \left(\frac{\frac{d}{dt} \phi_+(t)}{\phi_+(t)}\right), \label{decex}\\
\xi(t)&=&-2 {\rm Im}  \left(\frac{\frac{d}{dt} \phi_+(t)}{\phi_+(t)}\right)-2\omega_0,
\end{eqnarray}
and  $\{A,B\}$ denotes the anticommutator of $A$ and $B$. Physically speaking, the parameter $\kappa(t)$ represents the decay rate of the two-level impurity, while the renormalization parameter $\xi(t)$ plays the role of the Lamb shift due to the coupling to the lattice. 

The coupling of the impurity to the neighboring spins depends on the overlap between their wave functions, which fixes the magnitude of the exchange integral $g$. The overlap  depends on the size of the impurity  which could, for example, be an atom with one electron in the partially filled shell. Since the impurity is located halfway  between two lattice atoms  (in the center of a unit cell in general), the exchange between the latter and the impurity can be as strong as the  mutual coupling between the lattice constituents. In what follows, we  shall consider  the strong-coupling as well as the weak-coupling regimes of the dynamics.    

An example of the time dependence of the decay rate $\kappa(t)$ and the Lamb shift $\xi(t)$ is displayed in Figs.~\ref{figure2} and \ref{figure3} for some particular values of the model parameters. For convenience, the time and the magnetic field $h$ as well as $\omega_0$ are given in units of $J$.  It can be seen that  for strong coupling between the impurity and the lattice, the decay rate  takes on larger values  as the magnetic field increases  until the latter reaches  some  yet-to-be-determined critical value (which will be denoted from here on by $h_{\rm cri}$),  above  which the decay rate begins to decrease in magnitude. The Lamb  shift decreases in turn and after $h$ crosses its critical point the former becomes essentially negative. An other point worth observing is that  there appears  a peak which is followed by a sharp fall of the decay rate to negative values. This actually  happens at times of the order of $1/g$, which clearly is inversely proportional to the coupling constant.  The above results reveal the presence of a critical behavior of the dynamics with respect to the variation of the strength of the applied magnetic field.

It is worthwhile mentioning that the latter evolution features depend in a nontrivial way  on the impurity energy. The numerical investigation shows that they take place  in the strong-coupling regime ($g\sim J$) only when $\omega_0$ exceeds some threshold  value,  otherwise the decay rate always decreases as $h$ is raised. Nevertheless,  we only assign a critical value to the magnetic field, since, generally speaking, the latter is more accessible from an experimental point of view; this implies that we shall deal with $\omega_0$ as an intrinsic property of the impurity. 

\begin{figure}[htb]
{\centering{\resizebox*{0.42\textwidth}{!}{\includegraphics{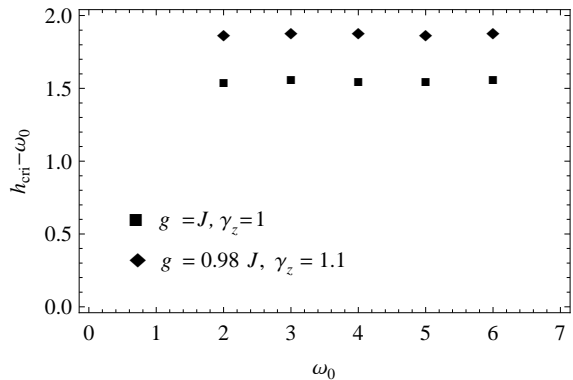}}}
\caption{\label{figure4} The numerical estimation of   $h_{\rm cri}$ as a function of $\omega_0$ for some values of $g$ and  $\gamma_z$ in the strong-coupling regime. The parameters $\omega_0$ and $h_{\rm cri}$ are given in units of $J$, and  $S=1$.}}
\end{figure}

Note that  as per the analytical expression of the impurity amplitude  $\phi_+(t)$, one cannot {\it a priori} determine the critical value  $h_{\rm cri}$ of the magnetic field. At first sight, it seems to be  in connection with the divergence of the  self-energy at the lower edge of the lattice spectrum.   We carried out many numerical calculations, and we always found that  $h_{\rm cri}< 4JS(1-\gamma_z)+\omega_0$. Actually, it may be put in  the form  $h_{\rm cri}= 4JS(1-\gamma_z)+\omega_0-\zeta(g)$ where $\zeta(g)$ is a positive monotonic  increasing function of $g$. In Fig.~\ref{figure4}, we display the numerical estimation of $\omega_0-h_{\rm cri}$ as a function of $\omega_0$ for some values of $g\sim J$. The near-constant outcomes suggest  the ansatz   $\zeta(g)=c g^2/ J$, where $c$ is a constant that is approximately equal to $3/2$. The latter yields a good fit to the numerical values, and may be used to locate the vicinity of the critical point for $g\sim J$. A more accurate fit gives $c=1.56$.  Hence we deduce that as far as the variation with respect to $h$ is concerned, the critical point always exists  when  $\omega_0>\zeta(g)-4JS(1-\gamma_z)$. Evidently, if  $\omega_0$ is very close to $\zeta(g)-4JS(1-\gamma_z)$, then  $h_{\rm cri}$ will also be close to zero, and its effect on the dynamics will not be so important, as the variation of the decay rate is quickly reversed by the increase of $h$.

 The above condition  explains the reason for which   the  critical  features of the dynamics  occur in the weak-coupling regime when $\gamma_z=1$  (i.e. Heisenberg lattice) practically for all values  of $\omega_0$, in contrast to the strong-coupling case. Indeed,  even for small $\omega_0$, there exists a value of $h$ above which the  decay rate always decreases, as was the case  in the strong-coupling regime for large $\omega_0$. Furthermore, for weak coupling, we notice the disappearance of the peak-shaped  variation of the decay rate and the Lamb shift. 
This regime is best investigated through a perturbative treatment;  the next section is devoted to these questions, which will be addressed in more detail when we derive the master equation within the second-order perturbation theory.

\begin{figure}[t]
{\centering\subfigure[{}]{\resizebox*{0.38\textwidth}{!}{\includegraphics{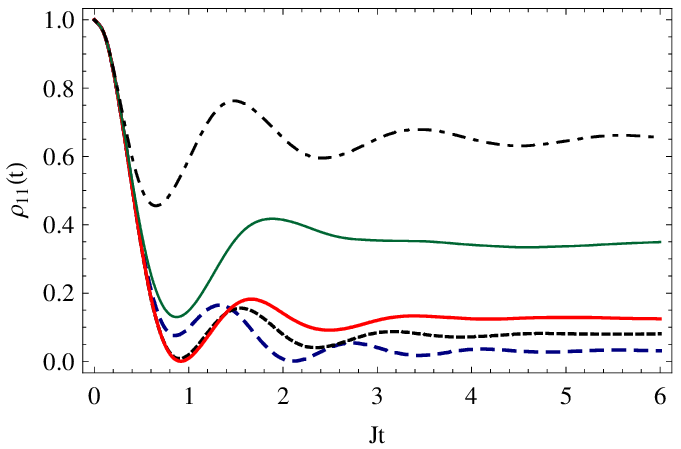}}\label{figure5-1}}
\subfigure[{}]{
\resizebox*{0.38\textwidth}{!}{\includegraphics{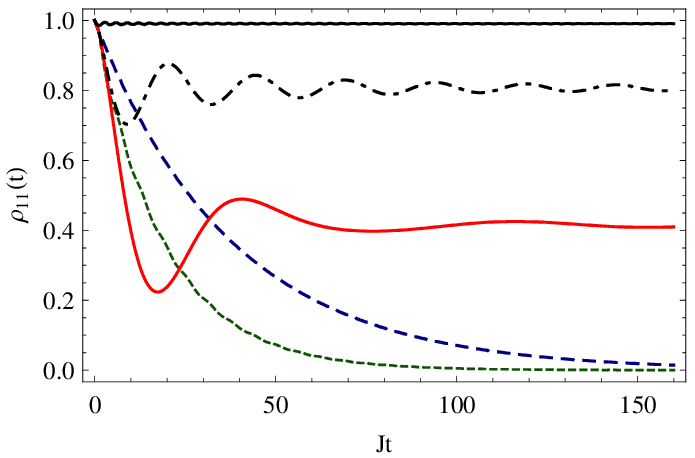}}\label{figure5-2}}
\par}
\caption{\label{figure5} The excited state occupation probability as a function of the time: (a) strong coupling with  $h=0.1J$ (blue dashed line), $h=J$ (black dotted line), $h=1.5J$ (thick red solid line), $h=3J$ (thin green solid line), and $h=4J$ (blue dot-dashed line); other parameters are: $\omega_0=3J$, $S=1$, $g=J$, $\gamma_z=1$. (b) Weak coupling with  $h=0.1J$ (blue dashed line), $h=2J$ (green dotted line), $h=3J$ (lower red solid line), $h=3.2 J$  (black dot-dashed line), and $h=5J$ (upper black solid line); other parameters are: $\omega_0=3J$,  $g=0.1J$,  $S=1$, $\gamma_z=1$. The impurity initially occupies the exited state, i.e. $\rho_{11}(0)=1$. }
\end{figure}

Now we turn to the investigation of the  evolution in  time of the reduced  density matrix of the impurity [see Fig. \ref{figure5}]. In accordance with the features exhibited by the decay rate, we find that for small values of $h$,  the matrix element  $\rho_{11}$, which represents the  occupation probability or population of the excited state, decreases  faster as the value of $h$ is raised, and mostly tends asymptotically to values very close to zero. The variation is reversed as we cross the critical point $ h_{\rm cri}$, and the occupation probability  decay becomes  slower; in particular the asymptotic state assumes larger values at  long times (see below for a quantitative discussion). For sufficiently strong magnetic field, the state of the impurity does not deviate much from its initial one. The time variation of the off-diagonal element $\rho_{12}$ exhibits essentially the same characteristics. This implies that decoherence of the state of the impurity may be minimized at moderate  times by applying a not too strong  (weak) magnetic field, but the asymptotic state at long times will be nearly diagonal; on the contrary,  if one is interested in the long-time behavior, it would be more convenient to apply a strong magnetic field. 

The increase of the magnetic field  should stabilize the ferromagnetic phase; this implies that, classically speaking, the lattice spins are more likely to be oriented along the $z$-direction. The effective strength of the $XY$ coupling should thus become weaker, leading to a lower decay rate. Our previous results,  however,  show that this is not the case at short to moderate intervals of time, but holds only at longer times. Specifically, the loss of coherence of the impurity and the decay of the occupation probability become more significant as we approach $h_{\rm cri}$ from bellow. 

 The observed sharp decrease of the decay rate can be accounted for as  the result of the fast  revival of  $\rho_{11}$ when $h$ is  close to $ h_{\rm cri}$; the revival is produced after the occupation probability has completely vanished. This is explained by the back-flow of information from the lattice to the impurity due to memory effects; these features correspond to the non-Markovian character of the dynamics, which holds even in  the weak-coupling regime. Specifically, we see from figure \ref{figure5-2} that for weak coupling, the near exponential decay of the density matrix element $\rho_{11}(t)$ is applicable only for small $h$; as we approach the critical point $h_{\rm cri}$, the decay becomes mostly Gaussian, and the asymptotic probability  does not vanish. A measure of the non-Markovianity of the dynamics  may be realized by investigating the sign of the decay rate. In either regime, whether weak or strong, the revival of  $\rho_{11}(t)$ corresponds to negative decay rates. Hence, we come to the important conclusion that even in the weak-coupling regime, the dynamics displays strong non-Markovian behavior. It may be approximated by the exponential (Markovian) law in the weak-coupling regime  only when the strength of the magnetic field is small enough, typically less than $h_{\rm cri}$. [see Sec.~\ref{sec4} for more details.] 
 
\begin{figure}[t]
{\centering\subfigure[{}]{\resizebox*{0.40\textwidth}{!}{\includegraphics{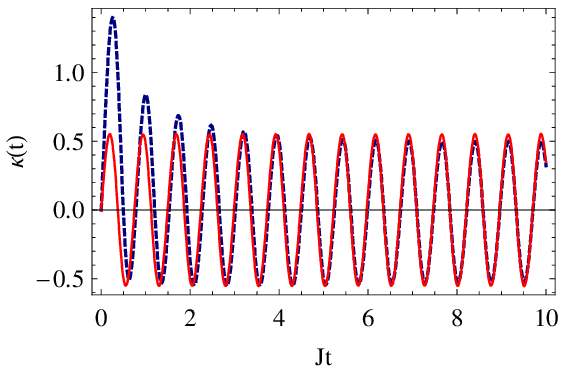}}}
\subfigure[{}]{
\resizebox*{0.40\textwidth}{!}{\includegraphics{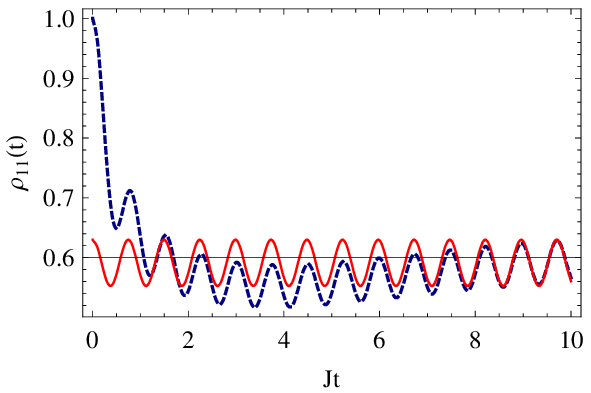}}}
\par}
\caption{ Asymptotic behavior of (a) the decay rate  $\kappa$, and (b) the excited state population $\rho_{11}(t)$  in the case of two bound states: The red solid lines  represent the asymptotic expressions of Eqs. (\ref{asym3}) and (\ref{asym4}), whereas the blue-dotted lines correspond to the exact solutions obtained by numerical integration of Eq.~(\ref{ampl}). The  parameters are $g=J$, $h=0.5J$, $\omega_0=8J$, $S=1$, and $\gamma_z=1$.  }\label{figure6}
\end{figure}

Analytically, the asymptotic value of $\phi_+$ can be determined by observing that by the Riemann-Lebesgue lemma,
\begin{equation} \medmath{
  \lim\limits_{t\to \infty}\int\limits_{\Omega_{\rm min}}^{\Omega_{\rm max}} \frac{\ {\rm Im}\ \Sigma_{\rm ret}(\epsilon+i\nu)e^{-i\epsilon t} d\epsilon}{\left(\epsilon-\omega_0-{\rm Re}\ \Sigma_{\rm ret}(\epsilon+i\nu)\right)^2+\left({\rm Im}\ \Sigma_{\rm ret}(\epsilon+i\nu)\right)^2}=0}.
\end{equation}
Therefore, we can distinguish between two cases: On the one hand, when the impurity possesses only one bound state, that is when $\omega_0\leq h+ 4JS(1+\gamma_z)h-g^2/J$,  whose energy is $\epsilon_1$, then:
\begin{eqnarray}
&&\lim\limits_{t\to \infty} |\phi_+(t)|^2=\frac{1}{\mathcal B(\epsilon_1)^2}, \label{asym1}\\
&& \lim\limits_{t\to \infty} {\kappa(t)}=0, \label{asym2}
\end{eqnarray}
where
\begin{equation}
\mathcal B(\epsilon)=1+\frac{4g^2}{(\epsilon-h-4JS(1-\gamma_z))[\frac{J}{g^2}(\omega_0-\epsilon)+1]}.
\end{equation}
The lamb shift $\xi(t)$ in turn tends to $2(\epsilon_1-\omega_0)$.  On the other hand, when the system exhibits two bound states, i.e. when $\omega_0> h+ 4JS(1+\gamma_z)h-g^2/J$, whose energies are $\epsilon_1<\epsilon_2$, then as $t\to\infty$:
\begin{eqnarray}
 |\phi_+(t)|^2&\sim&\frac{1}{\mathcal B(\epsilon_1)\mathcal B(\epsilon_2)}\left(\mathcal D (\epsilon_1,\epsilon_2)+2\cos[(\epsilon_2-\epsilon_1)t]\right),\label{asym3} \\
\kappa(t)&\sim& \frac{2(\epsilon_2-\epsilon_1)\sin[(\epsilon_2-\epsilon_1)t]}{\mathcal D(\epsilon_1,\epsilon_2)+2\cos[(\epsilon_2-\epsilon_2)t]},\label{asym4}
\end{eqnarray}
where
\begin{equation}
\mathcal D(\epsilon_1,\epsilon_2)=\frac{\mathcal B(\epsilon_1)}{\mathcal B(\epsilon_2)}+\frac{\mathcal B(\epsilon_2)}{\mathcal B(\epsilon_1)}.
\end{equation}
Hence, the asymptotic occupation probability oscillates in this case about $\mathcal D(\epsilon_1,\epsilon_2)/(\mathcal B(\epsilon_1)\mathcal B(\epsilon_2))$. The decay rate also displays periodic oscillation with amplitude inversely proportional to $\mathcal D(\epsilon_1,\epsilon_2)$. This is illustrated in Fig.\ref{figure6}. Notice that the effect of the anisotropy parameter $\gamma_z$ is merely to  renormalize the magnetic field $h$ in the low excitation sector of the Hamiltonian. Indeed, all the discussion presented thus far could be interpreted in terms of the effective field $\tilde{ h}=h+2\eta JS\gamma_z$. This is equivalent to shifting the critical points by the value $2\eta JS\gamma_z$. Hence, from here on, we shall focus mainly on the variation of the magnetic field $h$ and the impurity energy $\omega_0$. 
\subsection{Two-dimensional lattice}
 We now consider the  square lattice  in two dimensions, for which the lattice constants  in both the $x$ and the $y$ directions are the same and are equal to $\delta$.   Therefore,  the first Brillouin zone corresponds  to $-\pi/\delta\le k_i\le \pi/\delta$, $i\equiv x,y$.   We further  assume that the impurity lies in the center of a unit cell in the lattice so that the distance from it to any neighboring lattice spin is equal to $\delta/\sqrt{2}$. The coupling constant of the impurity to the lattice spins is denoted here also by $g$. Thus the squared modulus of the  coupling constant $g_{\vec k}$ is given by:
 \begin{eqnarray}
  |g_{\vec k}|^2&=&\frac{2S g^2}{N} \left|1+e^{i k_x \delta/2}(1+e^{i k_y \delta/2}+e^{i k_x \delta/2})\right|^2 \nonumber \\
  &=& \frac{32 g^2 S}{N} \cos^2(k_x\delta/2) \cos^2(k_y \delta/2).
 \end{eqnarray}
Moreover, the lattice structure factor reads now as:
\begin{equation}
 \tau_{\vec{k}}=\frac{1}{2}[\cos(k_x \delta)+\cos(k_y \delta)],
\end{equation}
whereas the spectrum bounds are $\Omega_{\rm min}= \tilde h-8JS$, and  $\Omega_{\rm max}= \tilde h+8JS$ (we use the notation $\tilde{ h}=h+8J S\gamma_z$). For large number of sites, the lattice spectrum turns into a continuum of states; in this limit, the retarded self-energy may be expressed as: 
\begin{widetext}
\begin{equation}
\Sigma_{\rm ret}(\epsilon+i\nu)=\frac{8g^2 S}{\pi^2}\int\limits_{-\pi}^{\pi}\int\limits_{-\pi}^{\pi}\frac{\cos^2(k_1/2) \cos^2(k_2/2) dk_1 dk_2}{\epsilon-h-8JS\gamma_z+4JS(\cos(k_1)+\cos(k_2))+i\nu}. \label{self2d}
\end{equation}
\end{widetext}
Outside the continuum, i.e.  $ \epsilon < \tilde h-8JS$ or $\epsilon > \tilde h+8JS$, the real part takes the form:
\begin{widetext}
\begin{eqnarray}
{\rm Re} \Sigma_{\rm ret}(\epsilon+i0^+)=\frac{g^2}{2\pi S J^2}\left[4\pi  S J -(\epsilon-\tilde h) E\left(\frac{64S^2J^2}{(\epsilon-\tilde h)^2}\right)+(\epsilon-\tilde h-8JS)K\left(\frac{64S^2J^2}{(\epsilon-\tilde h)^2}\right)\right], \label{in2d}
\end{eqnarray}
\end{widetext}
where $K$ and $E$ are the complete  elliptic integrals of the first and second kinds, respectively.

Inside the continuum, the integral in equation (\ref{self2d}) cannot directly be performed. Thus we analytically continue the right-hand side of equation (\ref{in2d}), by performing the analytic continuation of the complete elliptic integrals to the domain $|z|>1$, ${\rm Im}\  z<0$ of the complex plane, namely \cite{fettis}:
\begin{eqnarray}
K(z)&=&\frac{1}{\sqrt{z}}\left[K\left(\frac{1}{z}\right)-iK\left(1-\dfrac{1}{z}\right)\right],\\
E(z)&=&\sqrt{z} E\left(\frac{1}{z}\right)-\left(\frac{z-1}{\sqrt{z}}\right) K\left(\frac{1}{z}\right) \nonumber\\
&+& i\left[\sqrt{z} E\left(1-\dfrac{1}{z}\right)-\frac{1}{\sqrt{z}}K\left(1-\dfrac{1}{z}\right)\right].
\end{eqnarray}
 This yields
\begin{align}
{\rm Re} \Sigma_{\rm ret}(\epsilon+i0^+)=\frac{2g^2}{J}-\frac{4g^2}{J\pi} {\rm sgn} (\epsilon-\tilde h) E\left(\frac{(\epsilon-\tilde h)^2}{64J^2 S^2}\right)\nonumber \\ +\frac{g^2}{2S\pi J^2}\left[8JS\ {\rm sgn} (\epsilon-\tilde h)-|\epsilon-\tilde h|\right]K\left(\frac{(\epsilon-\tilde h)^2}{64J^2S^2}\right),
\end{align}
where ${\rm sgn}(x)$, denotes the sign of $x$. 
\begin{figure}[htb]
{\centering{\resizebox*{0.40\textwidth}{!}{\includegraphics{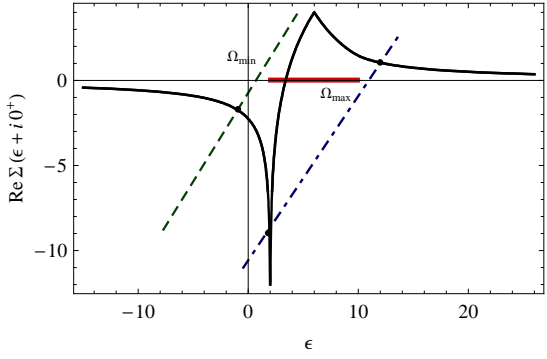}}}
\caption{\label{figure7}   The real part of the self-energy (black solid line) in a two-dimensional square lattice for $J=0.5$, $g=1$, $h=2$, $S=1$, and $\gamma_z=1$. The intersection points  correspond to the solutions of the equation ${\rm Re}\ \Sigma_{\rm ret}(\epsilon+i0^+)=\epsilon-\omega_0$ for $\omega_0=1$ (green dashed line) and $\omega_0=11$ (blue dot-dashed line); these solutions are interpreted in the main text as localized bound states outside the continuum.} }
\end{figure}
\begin{figure}[htb]
{\centering\subfigure[{}]{\resizebox*{0.38\textwidth}{!}{\includegraphics{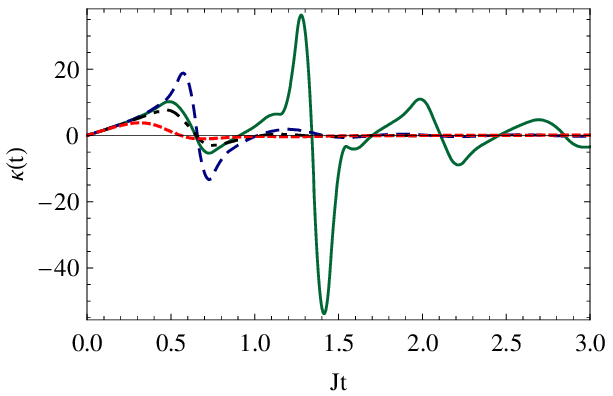}}}
\subfigure[{}]{
\resizebox*{0.38\textwidth}{!}{\includegraphics{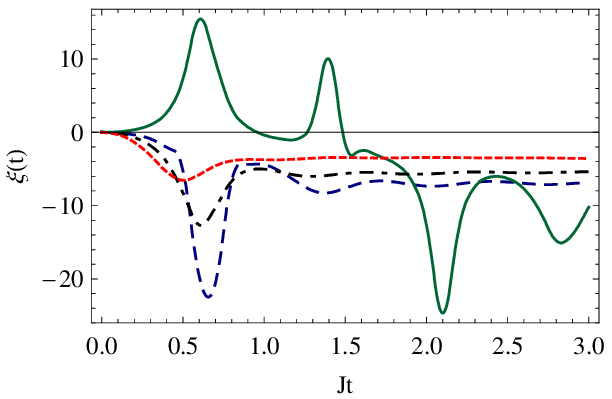}}}
\par}
\caption{\label{figure8} (a) The decay rate $\kappa$, and  (b) the Lamb shift $\xi$  for strong coupling in two dimensions as a function of the time for different values of the strength of the magnetic field: $h=0.1J$ (green solid lines), $h=2J$ (blue dashed lines), $h=3J$ (black dot-dashed lines), and $h=5J$ (red dotted line); other parameters are: $g=J$ $\omega_0=5J$, $S=1$, $\gamma_z=1$, $S=1$. }
\end{figure}

\begin{figure}[t]
{\centering\subfigure[{}]{\resizebox*{0.38\textwidth}{!}{\includegraphics{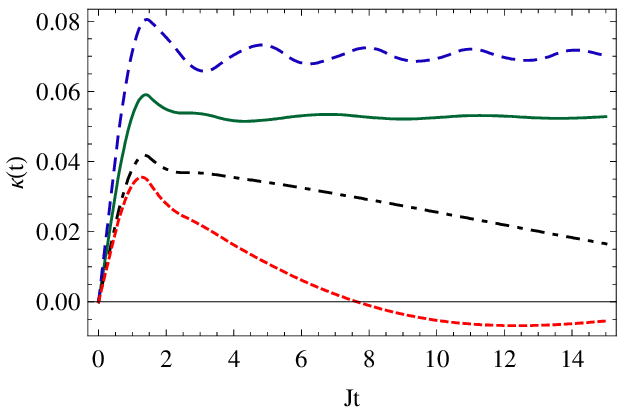}}}
\subfigure[{}]{
\resizebox*{0.38\textwidth}{!}{\includegraphics{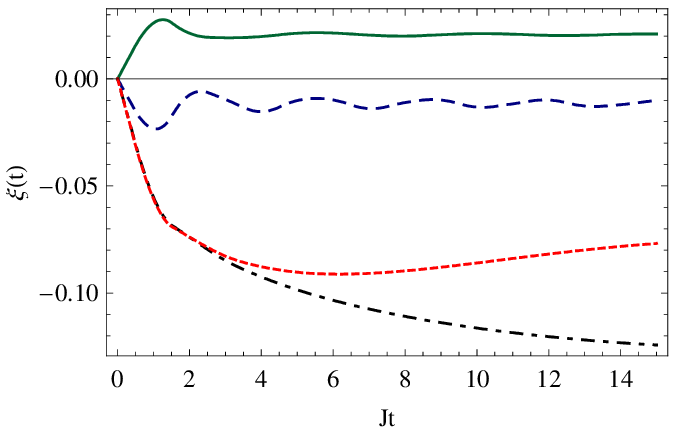}}}
\par}
\caption{\label{figure9} (a) The decay rate $\kappa$, and (b) the Lamb shift $\xi$  for weak coupling in two dimensions as a function of the time for different values of the strength of the magnetic field: $h=0.1J$ (green solid lines), $h=3J$ (blue dashed lines), $h=5J$ (black dot-dashed lines), and $h=5.2J$ (red dotted lines); other parameters are: $g=0.1J$, $\omega_0=5J$, $S=1$, $\gamma_z=1$, $S=1$.}
\end{figure}

Similarly, using the analytic continuation of $\Sigma(\epsilon)$, we find that outside the continuum ${\rm Im} \Sigma_{\rm ret}(\epsilon+i\nu)=0$, whereas inside the continuum, the imaginary part is calculated as:
\begin{eqnarray}
{\rm Im} \Sigma_{\rm ret}(\epsilon+i0^+&)=&\frac{g^2}{2\pi J }\left(\frac{\epsilon-\tilde h}{JS}\right)K\left(1-\frac{(\epsilon-\tilde h)^2}{64J^2S^2}\right)\nonumber \\&-&\frac{4g^2}{\pi J}E\left(1-\frac{(\epsilon-\tilde h)^2}{64J^2S^2}\right).
\end{eqnarray}
In this case, as depicted in figure~\ref{figure7},  it turns out that the system exhibits two bound states  when $\omega_0> h+8JS(1+\gamma_z)-{\rm Re}\Sigma_{\rm ret}(h+8JS(1+\gamma_z)+i0^+)$. In the opposite situation, there exists only one localized bound state. Moreover, we see that while in one dimension the real part of the  retarded self-energy of the impurity remains constant inside the continuum, in two dimensions, the same quantity diverges to negative values   above and bellow the lower bound of the lattice spectrum; in particular, it increases as we approach the upper bound where it takes on a finite value.   
\begin{figure}[htb]
{\centering\subfigure[{}]{\resizebox*{0.38\textwidth}{!}{\includegraphics{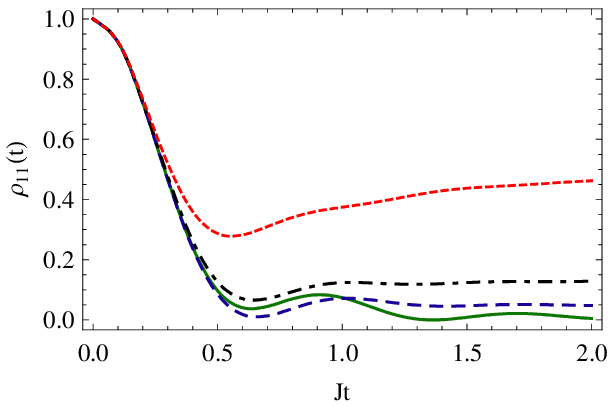}}}
\subfigure[{}]{
\resizebox*{0.38\textwidth}{!}{\includegraphics{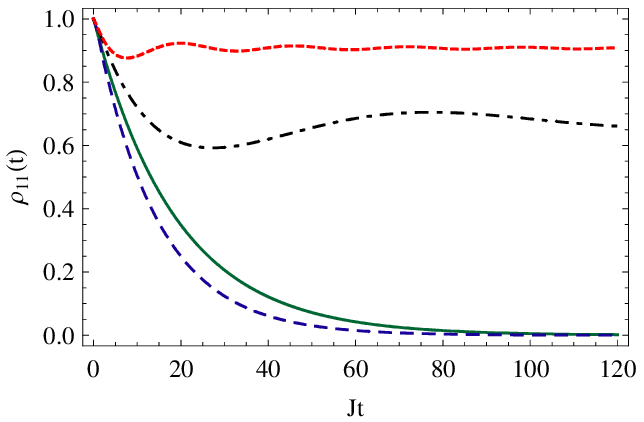}}}
\par}
\caption{\label{figure10}(color online)  The excited state occupation probability  as a function of the time in two dimensions. (a) Strong coupling with $h=0.1J$ (green solid line), $h=2J$ (blue dashed line), $h=3J$ (black dot-dashed line), and $h=5J$ (red dotted line); other parameters are: $g=J$, $\omega_0=5J$, $S=1$, and $\gamma_z=1$. (b) Weak coupling with  $h=0.1J$ (green solid line), $h=3J$ (blue dashed line), $h=5J$ (black dot-dashed line), and $h=5.2J$ (red doted line); other parameters are: $g=0.1J$, $\omega_0=5J$, $S=1$, and $\gamma_z=1$. The impurity initially occupies the exited state, i.e. $\rho_{11}(0)=1$.}
\end{figure}
The knowledge of the explicit form of the real and imaginary parts of the retarded self-energy makes it possible  to calculate  the amplitude $\phi_+(t)$ and the decay rate $\kappa(t)$. The obtained results are depicted in Figs.~\ref{figure8}-\ref{figure10} for both strong and weak coupling to the lattice. Here, again, it is found that the impurity dynamics is characterized by a critical dependence on  the applied magnetic field; all the results we have presented earlier in the case of the one-dimensional lattice hold in the two-dimensional one. The main difference rests in  the order of magnitude of the quantities of interest, which basically is due  to the increase of the number of nearest neighbors of the impurity.  The asymptotic values of the excited state occupation probability and the decay rate are given by expressions similar to equations ~(\ref{asym1}), (\ref{asym2}) for one bound state, and to equations~(\ref{asym3}), (\ref{asym4}) for two bound states, but in this case, we have:
\begin{eqnarray}
\mathcal B(\epsilon)&=&1+\frac{g^2}{2\pi J^2 S}\Bigl[\frac{\tilde h-\epsilon}{\tilde h-\epsilon-8J S}E\left(\frac{(\epsilon-\tilde h)^2}{64J^2S^2}\right) \nonumber \\
&-&\frac{\tilde h-\epsilon+8J S}{\tilde h-\epsilon}K\left(\frac{(\epsilon-\tilde h)^2}{64J^2S^2}\right)\Bigr].
\end{eqnarray}

It is interesting to notice that the features of the time evolution of the occupation probability, as obtained here, are quite similar to that of  Refs.~\cite{john,lambro} in photonic crystals displaying band gaps, where the critical behavior depends on the detuning from the band edge. The effect of the latter is thus equivalent to the effect of the  magnetic field in our spin system. While the oscillations in the photonic case  for large detuning are due to reflections from the dielectric host, the situation in the spin lattice has a different origin, namely the inhibition  of spin deviations in the lattice.   

\section{ Weak-coupling regime: perturbative treatment \label{sec4}}
The present section is devoted to the investigation of the dynamics  of the  impurity in the weak-coupling regime. This means
that the strength of the interaction  is taken sufficiently weak so to allow for a perturbative expansion with respect to the coupling constants $g_k$. For the sake of generality,  we assume that the lattice is in thermal equilibrium at temperature $T$, and that its state is initially uncorrelated from that of the impurity. This makes it easier to derive the evolution equations, since the zero-temperature dynamics is simply obtained by letting $T\to 0$. A discussion of the non-zero temperature case is given in the appendix~\ref{append}. This being said, it can now be shown that up to second order with respect to the coupling constants $g_{\vec k}$, the reduced density matrix verifies the master equation: 
\begin{eqnarray}
 \frac{d\rho(t)}{dt}&=&-[(\omega_0+\xi(t)-\xi^0(t)/2)\sigma_+\sigma_-,\rho(t)]\nonumber \\  &&\hspace*{-1.4cm}+\kappa(t)\left(\sigma_{-}\rho(t)\sigma_{+}-\frac{1}{2}\{\sigma_{+}\sigma_{-},\rho(t)\}\right)\nonumber \\
 &&\hspace*{-1.4cm}+(\kappa(t)-\kappa^0(t))\left(\sigma_{+}\rho\sigma_{-}-\frac{1}{2}\{\sigma_{-}\sigma_{+},\rho\}\right ), \label{mqs}
\end{eqnarray}
where $\kappa^0(t)=\kappa(t)|_{T=0}$, and $\xi^0(t)=\xi(t)|_{T=0}$. The decay rate and the Lamb shift at temperature $T$ are given by
\begin{equation}
\kappa(t)=2 \ {\rm Re} \Psi(t), \quad \xi(t)=2 \ {\rm Im} \Psi(t), \label{newpar}
\end{equation}
$\Psi(t)$ being the correlation function of the lattice, namely:
\begin{equation}
\Psi(t)=\sum_k  |g_{\vec k}|^2 e^{i(\omega_0-\Omega_{\vec k})t}[n(\Omega_{\vec k})+1]. \label{corr1} \\
  \end{equation}
In the above equation, $n(\Omega_k)$ denotes the mean number of magnons in mode $k$ at temperature $T$, that is:
\begin{equation}
 n(\Omega_{\vec k})=\frac{1}{e^{\Omega_{\vec k}/k_B T}-1}.
\end{equation}

 At zero temperature, the  master equation has the same form as the exact one (\ref{masteq}), and the solution at $T=0$ is thus given by (we drop the index):
\begin{eqnarray} 
 \rho_{11}(t)&=&\rho_{11}(0)\exp\left\{-\int_0^t\kappa(\tau)d\tau \right\},   \label{sol11}  \\ 
 \rho_{12}(t)&=& \rho_{12}(0) \exp\left\{-i\omega_0 t-\dfrac{i}{2}\int_0^t\xi(\tau)d\tau \right\}\nonumber \\&\times& \exp\left\{-\frac{1}{2}\int_0^t\kappa(\tau)d\tau \right\}.  \label{sol12}
\end{eqnarray}
This  form is quite general and is valid for both the exact and the second-order master equations. As a simple check, one can for instance insert $\kappa(t)$ as defined by equation (\ref{decex}) into equation (\ref{sol11}) to end up with the impurity amplitude.
\subsection{One-dimensional lattice}

The correlation function in the continuum limit  at zero temperature  takes the form (after a change of variable):

\begin{equation}
 \Psi(t)=4Sg^2 e^{i(\omega_0-h-4JS\gamma_z)t} \int\limits_{-1}^1 \frac{1+\zeta}{\pi\sqrt{1-\zeta^2}} e^{i4JS t \zeta} d\zeta\label{cor1}.
\end{equation}
This integral can be evaluated exactly  using the Bessel functions of the first kind, denoted here by ${\mathbf J}_n$,
yielding:
\begin{equation}
 \Psi(t)=4Sg^2 e^{i(\omega_0-h-4JS\gamma_z)t} \left[\mathbf J_0(4JSt)+i {\mathbf J}_1(4JSt)\right].
\end{equation}
Consequently, the decay rate  $\kappa$ can be expressed as [see Eq.~(\ref{newpar})]:
\begin{eqnarray}
 \kappa(t)=&8g^2 S & \int_0^t dt'\Bigl[\cos\bigl((\omega_0-\tilde h)t' \bigr) {\mathbf J}_0(4JSt')\nonumber \\ & -&\sin\bigl((\omega_0-\tilde h)t'\bigr){\mathbf J}_1(4JS t')\Bigr], \label{dec}
 \end{eqnarray}
 whereas the  Lamb-shift $\xi$ takes the form
 \begin{eqnarray}
 \xi(t)= & 8g^2 S & \int_0^t  dt'\Bigl[\sin\bigl((\omega_0-\tilde h)t'\bigr) {\mathbf J}_0(4JS t')\nonumber \\ & +&\cos\bigl((\omega_0-\tilde h) t'\bigr){\mathbf J}_1(4JS t')\Bigr].
\label{shif}	
\end{eqnarray}
\begin{figure}[t]
{\centering\subfigure[{}]{\resizebox*{0.38\textwidth}{!}{\includegraphics{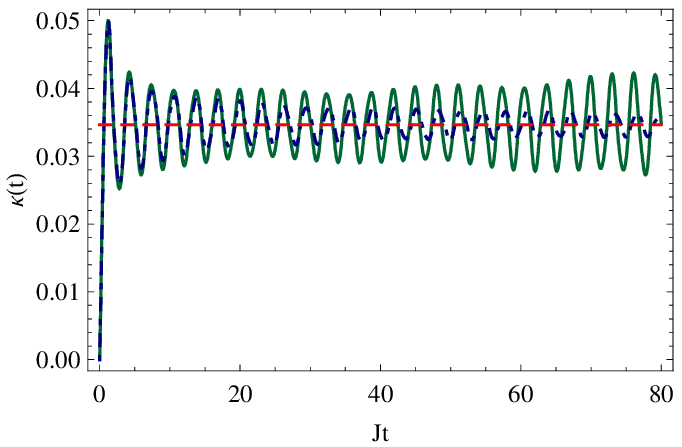}}}
\subfigure[{}]{
\resizebox*{0.38\textwidth}{!}{\includegraphics{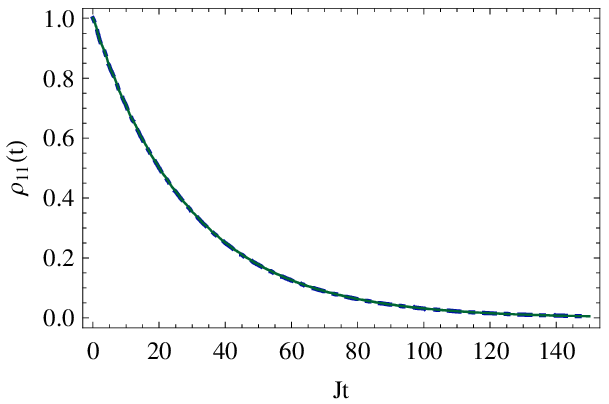}}}
\par}
\caption{\label{figure11}  Comparison between  the exact numerical solutions in one dimension (green solid lines) and  the outcomes of the second-order master equation (blue dot-dashed lines) in the weak-coupling regime for (a) the decay rate  and (b) the excited-state occupation probability; the parameters are $g=0.1 J$, $h=J$, $\omega_0=3J$, $S=1$,  $\gamma_z=1$, and $\rho_{11}(0)=1$; the horizontal red dashed line represents the Markovian decay rate. Notice that the two solutions for the occupation probability are almost identical for the chosen value of $h$.}
\end{figure}

In figure \ref{figure11}, we compare the decay rate  and the matrix element $\rho_{11}$ obtained here with the exact ones of Sec.~\ref{sec3}. It can be seen that the agreement is excellent for relatively long periods of time. In general, however, the two solutions do not coincide asymptotically, which is to be expected. In fact, the long-time behavior in this second-order approximation overestimates the actual exact values of the decay rate and the Lamb shift.
 Let us investigate the asymptotic values of the latter  quantities in the present approximation, which turn out to be given by  $\kappa_{\rm mark}=\lim_{t\to\infty}\kappa(t)=
           0$  for  $4JS<|\omega_0-h-4JS\gamma_z |$,  whereas
  \begin{equation}
   \kappa_{\rm mark}=  \frac{2g^2}{J}\sqrt{\frac{4JS-\omega_0+h+4JS\gamma_z}{\omega_0-h-4JS\gamma_z+4JS}}
  \end{equation}
   for $ 4JS>|\omega_0-h-4JS\gamma_z |$. Similarly, we find that $\xi_{\rm mark}=\lim_{t\to\infty}$ $ \xi(t)= \frac{2g^2}{J} $ for $ 4JS>|\omega_0-h-4JS\gamma_z| $ and 
   \begin{equation}
     \xi_{\rm mark}=   \frac{2g^2}{J}\left(1-\frac{|\omega_0-h-4JS(1+\gamma_z)|}{\sqrt{(\omega_0-h-4JS\gamma_z)^2-(4JS)^2}}\right) 
   \end{equation}
  for $ 4JS<|\omega_0-h-4JS\gamma_z| $, which are exactly the values of the decay rate and the Lamb shift obtained in the Markov approximation.
   \begin{figure}[t]
{\centering\subfigure[{}]{\resizebox*{0.38\textwidth}{!}{\includegraphics{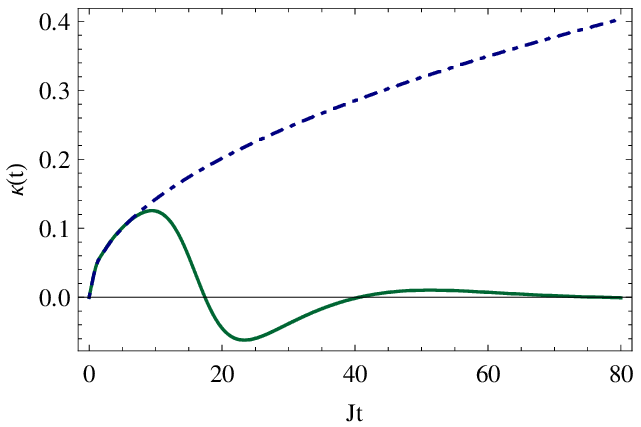}}}
\subfigure[{}]{
\resizebox*{0.38\textwidth}{!}{\includegraphics{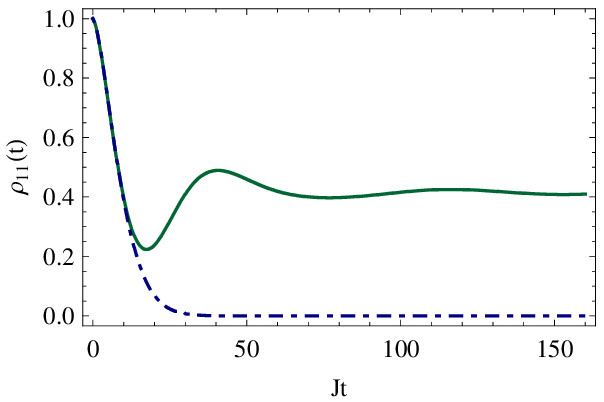}}}
\par}
\caption{\label{figure12}  The same as Fig.~\ref{figure11} but for $h=\omega_0=3J$ (resonance).}
\end{figure}
 The latter results are actually  a manifestation of the  breakdown of the Fermi golden rule which states for instance that, for weak coupling, the decay rate and the Lamb shift  are given by  $\kappa_{\rm mark}=2\ {\rm Im}\Sigma_{\rm ret}( \omega_0+i0^+)$, and   $\xi_{\rm mark}=2\ {\rm Re}\Sigma_{\rm ret}( \omega_0+i0^+)$.
 
 We have already noticed in Sec.~\ref{sec3} that, in the weak-coupling regime, the Markovian  decay  law $e^{-\kappa_{\rm mark}t} $ is valid only for weak magnetic field. The reason behind this resides in the fact that the decay of the correlation function of the lattice is fast enough only when $h$  is small. The larger the values of $h$,  the slower the decay of the correlation function is; the latter exhibits in particular oscillatory variation at long times, and hence   the dynamics deviates from the exponential law to the Gaussian one; in all cases, the long-time limit of the quantities $\kappa$ and $\xi$ exists  thanks to the properties of the Bessel functions of the first kind. The decay of the correlation function is a general property of the weak coupling to continua, which is the case in this model. The breakdown of the Fermi golden rule is best illustrated by the vanishing value of the decay rate   $\kappa_{\rm mark}=0$  for  $4JS<|\omega_0-h-4JS\gamma_z |$. Hence, if we apply directly the latter rule, we find that there occurs no decay of the state of the impurity; in other words, the impurity does not feel at all the presence of the lattice despite its coupling to the latter, which  is  not necessarily the case as is confirmed by the exact solution of the previous section.  It is also worthwhile noticing that although the Markovian limit fails to reproduce the actual dynamics  at long times, it keeps track of the overall critical behavior of the impurity as is discussed bellow.

\subsubsection*{ Resonance-like behavior}
A particular instance  occurs  when $\omega_0$ coincides with $\Omega_k$ in the center or at the edges of the first Brillouin zone, that is for $k=0, \pm \pi$.  In this resonance-like case, we can distinguish two possible situations, namely $\omega_0-h-4J S\gamma_z=\pm4JS$. We begin with the condition $\omega_0= h+4JS(\gamma_z-1)$, which  should  be compared with the one obtained in Sec.~{\ref{sec3} for $h_{\rm cri}$, evaluated in the weak-coupling regime $g\ll J$, i.e.: $ h_{\rm cri}=4J S(1-\gamma_z)+\omega_0$.  The latter relation is a very peculiar condition that links the energy of the two level impurity to the lower limit of the spectrum of the lattice. It  occurs precisely in  the center of the first Brillouin zone. The particular feature of the decay rate and the Lamb shift in this case rests in the fact that they grow relatively fast as the time increases; in particular, we find that the Markovian limits diverge since $\lim_{t\to\infty}\kappa(t)=\lim_{t\to\infty}|\xi(t)|=\infty$. A comparison between the exact decay rate and the perturbative one is carried out in Fig. \ref{figure12},  where we can see that initially, the two coincide at short times, but eventually the exact decay rate tends asymptotically to zero.  We also notice that while the approximate decay rate  remains positive, the exact one takes negative values, indicating regeneration of both  the excited state occupation probability and the quantum interferences (recoherence).
On the other hand, when $\omega_0= h+4JS(1+\gamma_z)$ we obtain a rather reduced decay rate, and in particular, it turns out that $\kappa(t)\to 0 $ while $\xi(t)\to   2g^2/J$ as $t \to\infty$; this indicates that there occurs no divergence of the decay rate and the Lamb shift in the Markov approximation. 
 
 It should be noted that the condition $4JS\le|\omega_0-h-4JS\gamma_z|$ is equivalent to the statement that $\omega_0$, the characteristic intrinsic energy level-spacing  of the impurity, lies within the continuum associated with the lattice. The  divergence of the Markovian decay rate may typically be  attributed   to a resonance  in the center of the first Brillouin zone where  $\omega_0=h+4JS(\gamma_z-1)$.  This can be explained by  the coupling of the impurity to the  collective mode-zero of the spin degrees of freedom of the lattice. In this mode, the effects of the quantum excitations or magnons add to each other coherently, and hence it dominates over the other modes. Indeed,  by inspecting equation (\ref{cor1}), we see that the   spectral function is given by $f(z)=(1+z)/\pi{(1-z^2)^{1/2}}$, which clearly displays a Van Hove singularity only in the center of the first Brillouin zone, i.e. when $z=1$ or equivalently  $k=0$, and vanishes at its edges where $z=-1$,  which corresponds to $k=\pm\pi$. Once the parameter $\omega_0$  exceeds the lower bound, the decay rate begins to decrease as the former approaches the upper bound of the spectrum. The above variation persists even when $\omega_0$ exits the continuum. When $\omega_0<h$, the resonance condition cannot be satisfied, which explains the absence of the critical divergence in the Markov limit.

 \subsection{Two and three-dimensional lattices}

The decay  of the correlation function  in two and three-dimensional lattices is much faster than that of the one-dimensional lattice even for small $h$; the suppression of the oscillations is more noticeable at shorter times.  The exact decay rate and the excited-state occupation probability  $\rho_{11}$, along with the approximate ones are illustrated in Fig.~\ref{figure13} for $d=2$. The agreement is good for long times.  Furthermore, applying Fermi's golden rule, the Markovian decay rate in the case of the square lattice vanishes for  $|\omega_0-\tilde h|>8JS$; when $|\omega_0-\tilde h|<8JS$ it is given by
\begin{eqnarray}
\kappa_{\rm mark}&=&-\frac{g^2}{\pi J }\left(\frac{\omega_0-\tilde h}{JS}\right)K\left(1-\frac{(\omega_0-\tilde h)^2}{64J^2S^2}\right)\nonumber \\&+&\frac{8g^2}{\pi J}E\left(1-\frac{(\omega_0-\tilde h)^2}{64J^2S^2}\right).
\end{eqnarray}
At resonance, $h-\omega_0=8JS(1-\gamma_z)$, the Markovian decay rate remains finite; indeed, on account of the fact that $K(0)=E(0)=\frac{\pi}{2}$, it follows that 
\begin{equation}
\kappa_{\rm mark}=\frac{4g^2}{J}.
\end{equation}
If the impurity possesses one bound state, the exact decay rate vanishes at infinity, and hence it differs significantly from the Markovian rate, see figure \ref{figure14}.

The coupling constant in a three-dimensional simple cubic lattice, where the impurity occupies the center of a unit cell,  is given by:
\begin{eqnarray}
g_{\vec k}&=&g\sqrt{\frac{2S}{N}}\Bigl[\cos[\frac{\delta}{2}(k_x+k_y+k_z)]+\cos[\frac{\delta}{2}(k_x-k_y+k_z)]\nonumber \\ &+&\cos[\frac{\delta}{2}(k_x+k_y-k_z)]+\cos[\frac{\delta}{2}(-k_x+k_y+k_z)]\Bigr].
\end{eqnarray}
It follows  that:
\begin{equation}
|g_{\vec k}|^2=\frac{128g^2 S}{N} \cos(\delta k_x/2)^2 \cos(\delta k_y/2)^2 \cos(\delta k_z/2)^2.
\end{equation}
The integration with respect to the wave vector $\vec{k}$ in the continuum limit is more involved here, but we can nevertheless draw the following conclusion: 
The density of states of the lattice is finite; therefore, the decay rate and the Lamb shift in the Markovian limit do not diverge, even at resonance, as illustrated in Fig.\ref{figure15}. Here, also, the time evolution depends on whether $h$ exceeds  the critical value, which turns out to be $h_{\rm cri}=\omega_0+12JS(1-\gamma_z)$.
\begin{figure}[t]
{\centering\subfigure[{}]{\resizebox*{0.38\textwidth}{!}{\includegraphics{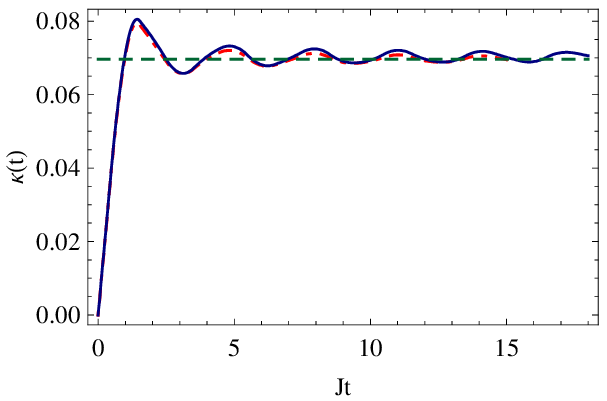}}}
\subfigure[{}]{
\resizebox*{0.38\textwidth}{!}{\includegraphics{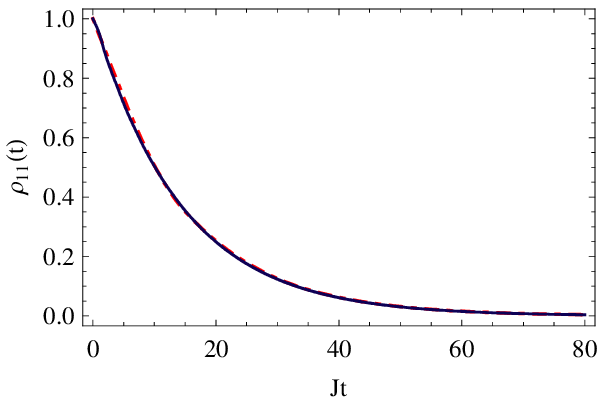}}}
\par}
\caption{\label{figure13}  Time evolution of  (a) the decay rate $\kappa$ and  (b) the excited state occupation probability $\rho_{11}$  in two dimensions at resonance with $h=\omega_0=5J$ for weak coupling of the impurity: exact solution (blue solid lines), and the  solution of the second-order master equation (red dot-dashed lines). The other parameters are: $g=0.1 J$, $S=1$, and $\gamma_z=1$. Notice the complete suppression of oscillations at long times in two dimensions.}
\end{figure}
  
\begin{figure}[htb]
{\centering{\resizebox*{0.38\textwidth}{!}{\includegraphics{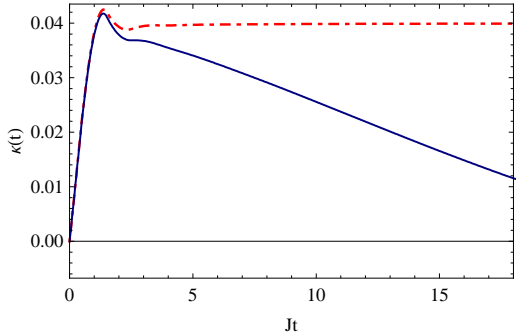}}}
\caption{\label{figure14} The  decay rate   in two dimensions as a function of the time: exact solution (blue solid line), and the second-order approximation (red dot-dashed line). The parameters are   $h=3J$,  $\omega_0=5J$, $g=0.1 J$, $S=1$, and $\gamma
=1$.}}
\end{figure}
  \begin{figure}[htb]
{\centering{\resizebox*{0.38\textwidth}{!}{\includegraphics{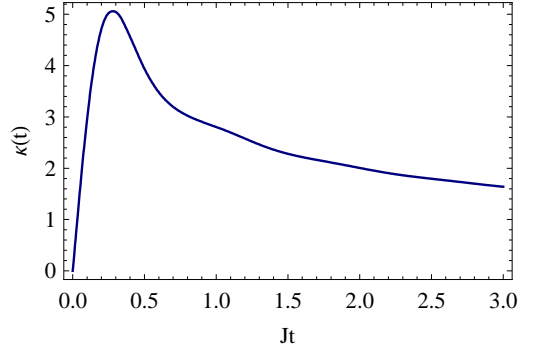}}}
\caption{\label{figure15}  The approximate decay rate in three dimensions at resonance with $h=\omega_0=3J$ for weak coupling of the impurity. The other parameters are: $g=0.1 J$, $S=1$, and $\gamma_z=1$.}}
\end{figure}

\subsection{Short-time variation}
From the above discussion, one can conclude that in a $d$-dimensional lattice, we have in general:
\begin{equation}
 |g_{\vec k}|^2=\frac{2^{2d+1}g^2 S}{N} \prod_{i=x,y,z}\cos(\delta k_i/2)^2.
 \end{equation}
By expanding the integrand of Eq.~(\ref{dec}) in Taylor series, and integrating term by term,  it follows that in the continuum limit, the variation of the decay rate  at short times is described by:
\begin{equation}
 \kappa(t)=2^{d+2} g^2 S t+O(t^3)\label{decshort},
\end{equation}
which is linear in  time and is independent of the magnetic field. This, however, is not the case for the Lamb shift which is is affected by  both the magnetic field and the impurity energy. For instance, in $d=1$, it turns out that
\begin{equation}
  \xi(t)=4g^2 S (\omega_0-h+2JS(1-2\gamma_z)) t^2+O(t^4).
 \end{equation}
  Therefore, if $h+4JS(\gamma_z-1/2)>\omega_0$, the coefficient of $t^2$ becomes negative,  in  complete accordance with the observed decrease of the Lamb shift to negative values.
  
Taking into account equations (\ref{sol11}) and  (\ref{sol12}), we obtain that at short times:
\begin{eqnarray}
\rho_{11}(t)&\simeq& \rho_{11}(0) e^{-2 t^2/\tau_D^2}, \label{short1}\\
 |\rho_{12}(t)|&\simeq& |\rho_{12}(0)| e^{- t^2/\tau_D^2},\label{short2}
\end{eqnarray}
where the decoherence time constant $\tau_D$ is defined by:
\begin{equation}
\tau_D=\frac{1}{2^{d/2} g\sqrt{S}}.\label{constdec}
\end{equation}
These expressions are best applied to the strong-coupling regime; they turn out to be a very good approximation in particular for small values of $h$, as is illustrated in Fig.\ref{figure16}, where we display the exact evolution in time of the density matrix element $\rho_{11}(t)$ along with the approximate one given by equation~(\ref{short1}). 

The linear dependence of the decay rate, together with the quadratic behavior of the decay of the reduced density matrix  at short times are known to correspond to the so-called Zeno regime~\cite{misra, home, itano, home2, fischer, streed, silva,wu,debi,lern,facchi2, facchi,kof,zhang3}. The investigation of this regime in the context of the present work makes the subject of the next section.

\begin{figure}[htb]
{\centering{\resizebox*{0.38\textwidth}{!}{\includegraphics{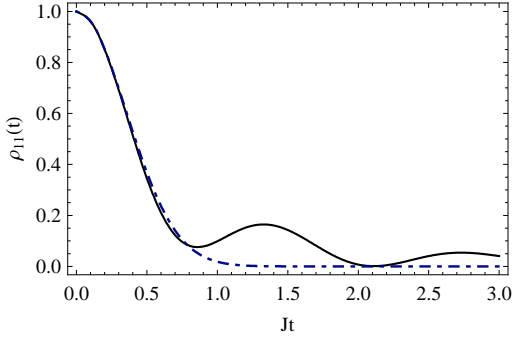}}}
\caption{\label{figure16}   Comparison between the exact numerical solution in one dimension (black solid line)  and the short-time approximation  (blue dot-dashed line)  of Eq.~(\ref{short1}). The parameters are $g=J$, $h=0.1J$, $\omega_0=3J$, $S=1$,  $\gamma_z=1$, and $\rho_{11}(0)=1.$ }}
\end{figure}
 
 \section{Application to the Quantum Zeno effect (QZE)\label{sec5}}
Let us begin by recalling the main ideas behind the concept of the quantum Zeno effect as applied to the impurity \cite{facchi}.  Suppose that the latter is initially prepared  in the excited state, which is equivalent to setting $\rho_{11}(0)=1$. As the time evolves, the so-called survival probability is given by $P(t)=\rho_{11}(t)$. This is the probability of finding the impurity at later times in the initial state. If a series of $N$ measurements are performed at regular time intervals $\tau$, the survival probability becomes 
\begin{equation}
P(N\tau)=P(\tau)^N=\rho_{11}(\tau)^N.
\end{equation}
An effective decay rate is introduced  via the identity:
\begin{equation}
P(N\tau)=e^{-\kappa_{\rm eff}(\tau) t},
\end{equation}
where $t=N\tau$. Notice that by Eq.~(\ref{genden}), we have $\rho_{11}(t)=|\phi_+(t)|^2$; it immediately follows that:
\begin{equation}
\kappa_{\rm eff}(\tau) =-\frac{1}{\tau}\ln(\rho_{11}(\tau))=-\frac{1}{\tau}\ln|\phi_+(\tau)|^2,
\end{equation}
which should be compared with the exact decay rate of equation (\ref{decex}) that can be written  as:
\begin{equation}
\kappa(t)=-\frac{d \ln|\phi_+(t)|^2}{dt}.
\end{equation}
The two decay rates are generally different as illustrated in figure \ref{figure17}; in fact, even at short times, the above expressions yield distinct outcomes. For instance,  consider the  weak-coupling regime  which is described by equations~(\ref{corr1}) and~(\ref{sol11}) at $T=0$; these give (for ease of notation we drop the vector symbol) :
\begin{eqnarray}
\int\limits_0^t \kappa(t') dt'&=&2\sum_k |g_k|^2\int\limits_0^t\frac{ \sin((\omega_0-\Omega_k)t')}{\omega_0-\Omega_k}dt'\nonumber \\
&=& 2 t^2 \sum_k |g_k|^2 \frac{ \sin^2((\omega_0-\Omega_k)t/2)}{[(\omega_0-\Omega_k) t/2]^2}.
\end{eqnarray}
Therefore, for a measurement time $\tau$: 
\begin{equation}
\int_0^\tau \kappa(t')dt'=2\tau \kappa_{\rm eff}(\tau),
\end{equation}
 where \cite{facchi,kof,zhang3}
\begin{equation}
 \kappa_{\rm eff}(\tau)=\tau \sum_k |g_k|^2 \frac{ \sin^2((\omega_0-\Omega_k)\tau/2)}{[(\omega_0-\Omega_k) \tau/2]^2}.
 \end{equation}
 For small measurement time, we have:
\begin{equation}
\int\limits_0^\tau \kappa(t') dt'\simeq [\kappa(\tau)-\kappa(0)] \tau=\tau \kappa(\tau),
\end{equation}
which shows that the effective decay rate is twice smaller than the exact decay rate. In fact if we keep only terms linear in $\tau$ in the expansion of the sine function, we end up with
 \begin{equation}
\kappa_{\rm eff}(\tau)=2^{d+1} g^2 S\tau=\frac{\kappa(\tau)}{2},\label{zen1}
\end{equation}
in complete agreement with equation (\ref{decshort})  describing the short time variation at zero temperature.
\begin{figure}[htb]
{\centering{\resizebox*{0.38\textwidth}{!}{\includegraphics{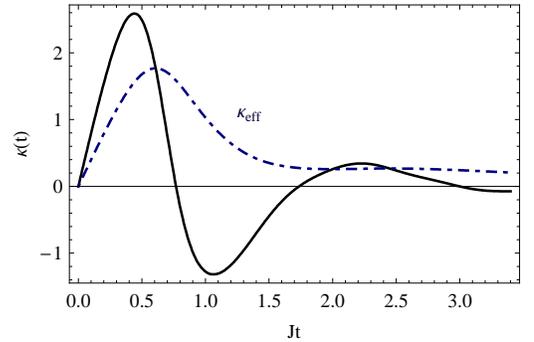}}}
\caption{\label{figure17}  The exact (black solid line) and the effective (blue dot-dashed line) decay rates in one dimension as a function of time for $g=J$, $h=3J$, $\omega_0=2J$, $S=1$, and $\gamma_z=1$.}}
\end{figure}
Notice, moreover,  that the effective decay rate remains always positive at all times, which is obvious from its definition because $|\phi_+(\tau)|^2\le 1$. This means that for large  $\tau$, the effective decay rate is insensitive to the regions of negative exact decay rate, which correspond to recoherence effects as we have  mentioned earlier. From the above results, we see that one obtains always the QZE at short times, as the measurement slows down the decay of the impurity. At larger times, however, one may thus obtain the  quantum inverse Zeno effect (IZE)~\cite{zhang3}.   The recoherence effects become more appreciable for large values of the magnetic field; for such values, the measurement may have a destructive effect on the coherences of the impurity, and may thus also lead to the acceleration of the decay of the survival probability. 
 
 More importantly, it becomes clear  from equation~(\ref{zen1}) that, if the measurement is performed at time scales for which the latter equation is valid, the impurity evolution becomes  independent of both the magnetic field $h$ and the intrinsic level energy-spacing $\omega_0$. In particular, when $h>h_{\rm cri}$ the measurement speeds up the decay of the survival probability, preventing thus the magnetic field  from protecting the impurity state from the effect of the lattice, which is a direct consequence of the IZE. 
 
 Let us now assume that  $h$ is small enough so that we ascertain that the Fermi golden rule holds for weak coupling. Under these conditions, the Markovian decay rate defines the natural life time $1/\kappa_{\rm mark}$ of the impurity. In this case,  according to Ref.~\cite{facchi}, the criterion for the QZE to happen is that $\tau$ be smaller than $\tau^*$, the solution of the equation $\kappa_{\rm eff}(\tau^*)=\kappa_{\rm mark}$. When $\tau>\tau^*$, the IZE takes place. We have seen that for $\omega_0<h-2J\eta S(1-\gamma_z)$ or $\omega_0>h+2J\eta S(1+\gamma_z)$, the Markovian decay rate vanishes, i.e.  $\kappa_{\rm mark}=0$. In this case there exists no solution for the latter equation, and we obtain always the IZE.
\section{Conclusion}
The present study gives a thorough discussion of the dynamics of a two-level  impurity that is coupled through XY interaction to a ferromagnetic lattice at low temperatures.  Under the condition of small lattice excitations, our model is equivalent to the Fano-Anderson one, with a particular form of the coupling constant, which  is due to the geometric configuration of the system where the impurity occupies the center of a unit cell in the lattice. This makes it possible to derive in an exact manner the zero-temperature retarded Green's function in one and two dimensions. The latter is directly linked to the excited state  amplitude, which is found to satisfy a master equation in Lindbald form involving the decay rate and the Lamb shift. By studying the evolution of those  quantities, we find that under certain conditions, there exits a critical value of the magnetic field above which the decay always slows down. In the weak-coupling regime, the critical point occurs when the impurity energy coincides with the lower bound of the continuum. In particular, in the case of the Heisenberg model, for which the anisotropy parameter $\gamma_z$ is set to unity, the  critical magnetic field is identical to the impurity level energy-spacing, which we termed resonance. The investigation reveals that in this regime, the Fermi golden rule does not apply if the magnetic field exceeds the critical value. The exponential decay law holds only for weak magnetic fields, for which the lattice correlation function is damped fast enough so that the conditions of the Markovian approximation are fulfilled. We have derived the master equation  for the reduced density matrix of the purity in the weak-coupling regime. The elimination of the lattice degrees of freedom is carried out by taking into account the spectral properties of the lattice which are uniquely fixed by its dispersion relation. The validity of the master equation is discussed by comparing its outcome with the exact solution. At resonance,  the Markovian decay rate and the Lamb shift diverge in one dimension, but remain finite at higher dimensions. The effective  decay rate of the Zeno effect is found to be insensitive to regions of negative decay rate, and hence the measurement may lead to the inverse Zeno effect, as the decay may be accelerated, in particular for strong magnetic field. 
 \acknowledgments
The author would like to thank the referee for the valuable suggestions and comments.

\appendix*

  \section{Effect of the temperature} \label{append}

 The spin-wave formalism is  applicable at low temperatures, where the number of magnons or excitations is small. The main criterion for the use of the Holstein-Primakoff transformation is $
 n(\Omega_k)\ll 2S$ for all modes (we drop the vector symbol).  Actually, it is sufficient that the lower bound  of the lattice spectrum $\Omega_{\rm min}=h+2J S\eta(\gamma_z-1)$ verifies the above criterion to ensure that the mean numbers of magnons in all modes are small enough,  which can be formulated as:
\begin{equation}
 (e^{ [h+2 S J\eta(\gamma_z-1)]/k_BT}-1)^{-1} \ll 2S.
\end{equation}
For temperatures satisfying the latter condition, the density matrix in the weak-coupling regime is described by the master equation~(\ref{mqs}). It is a matter of algebra to show that its solution is given by:
 \begin{align} 
   \rho_{11}(t)=\exp\left\{-\int_0^t(2\kappa(\tau)-\kappa^0(\tau))d\tau \right\}  \Bigl[\rho_{11}(0) + \nonumber \\\int_0^t (\kappa(\tau)-\kappa^0(\tau)) \exp \Bigl\{\int_0^\tau(2\kappa(\tau')-\kappa^0(\tau'))d\tau'\Bigr \}d\tau\Bigr],\\
   \rho_{12}(t)= \rho_{12}(0) \exp\left\{-i\omega_0t-\dfrac{i}{2}\int_0^t(2\xi(\tau)-\xi^0(\tau))d\tau \right\}\nonumber \\ \times
     \exp\left\{-\frac{1}{2}\int_0^t (2\kappa(\tau)-\kappa^0(\tau) )d\tau \right\}.
 \end{align}
 \begin{figure}[t]
{\centering\subfigure[{}]{\resizebox*{0.34\textwidth}{!}{\includegraphics{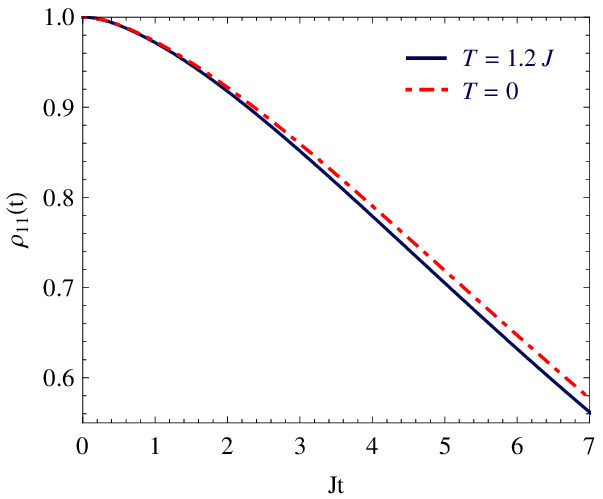}}\label{figure18-1}}
\subfigure[{}]{
\resizebox*{0.34\textwidth}{!}{\includegraphics{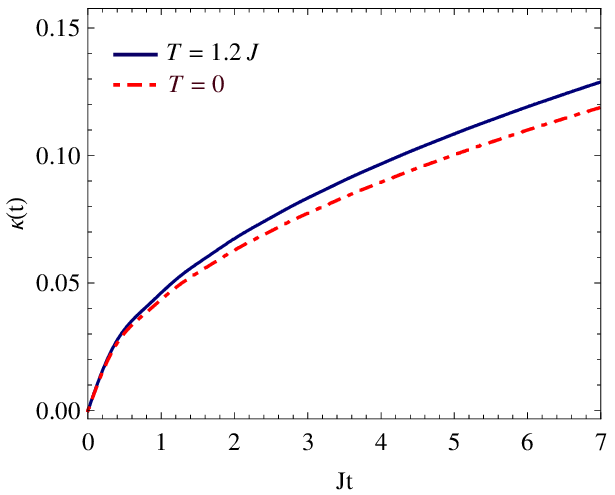}}\label{figure18-2}}
\par}
\caption{\label{figure18}  The time dependence of (a)  the excited state population $\rho_{11}$, and (b) the decay rate $\kappa$ at zero temperature (red dot-dashed lines) and non-zero temperature (blue solid lines) in the weak-coupling regime at resonance with $g=0.1J$,  $h=\omega_0=3J$, $S=1$, $\gamma_z=1$, and $\rho_{11}(0)=1$ (we set $k_B=1$).}
\end{figure}
 
  Figure \ref{figure18} gives an example of the time variation of the decay rate  at nonzero temperature in one dimension. It can be seen that as expected, the decay rate becomes larger as the temperature raises, which is due to the fact that the number of magnons becomes more important, leading to greater deviations of the spin vectors toward the $x$-$y$ plane; as a result,  the effective $XY$ coupling of the impurity to the lattice also grows.  At such low temperatures, the critical dependence of the decay rate on the magnetic field still holds, which means that the Markovian decay rate diverges when $h=\omega_0+4JS(1-\gamma_z)$.
  
 At sufficiently short times, we may approximate the reduced density matrix elements by:
 \begin{eqnarray}
 \rho_{11}(t)&=&e^{-2\Gamma t^2}\left(\rho_{11}(0)-\frac{\Delta}{\Gamma}\right)+\frac{\Delta}{\Gamma},\\
  |\rho_{12}(t)|&=& |\rho_{12}(0)|e^{-\Gamma t^2},
 \end{eqnarray}
 where 
 \begin{eqnarray}
 \Gamma&=&\frac{1}{2}\sum_{ k} |g_{ k}|^2 (2n(\Omega_{ k})+1),\\
 \Delta&=&\frac{1}{2}\sum_{ k} |g_{ k}|^2 n(\Omega_{ k}).
 \end{eqnarray}
 It follows that the temperature-dependent decoherence time constant is given by
 \begin{equation}
 \tau_D=\frac{\sqrt{2}}{\sqrt{\sum_{ k} |g_{ k}|^2 (2n(\Omega_{ k})+1)}}.
 \end{equation}

\end{document}